\documentclass[superscriptaddress,aps,pra,twocolumn,showpacs,nofootinbib,longbibliography,notitlepage,10pt]{revtex4-2}
\usepackage{etex}
\usepackage{Physics}
\usepackage{amsmath,amssymb,amsthm,mathrsfs}
\usepackage[colorlinks=true,citecolor=blue,urlcolor=blue]{hyperref}
\usepackage{subfigure}
\usepackage[pdftex]{graphicx}
\usepackage{times,txfonts}
\usepackage{braket}
\usepackage{color}
\usepackage{natbib}
\usepackage{amsmath,blkarray}
\usepackage{mathtools}
\usepackage{latexsym}
\usepackage{tabularx, booktabs}
\usepackage{graphics,epstopdf}
\usepackage{graphicx}
\usepackage{float}
\usepackage{graphicx}
\usepackage{amsfonts}
\usepackage{color,soul}

\begin{document}

	\title{Nonlocal correlations in the asymmetric quantum network}
 \author{Souradeep Sasmal}
	\email{souradeep.007@gmail.com}
 \affiliation{National Institute of Technology Patna, Ashok Rajpath, Patna, Bihar 800005, India}
 \affiliation{Department of Physics, Indian Institute of Technology Hyderabad, Telengana-502284, India }
	\author{ Shyam Sundar Mahato }
	\email{shyamsundarmahato59@gmail.com}
 \affiliation{National Institute of Technology Patna, Ashok Rajpath, Patna, Bihar 800005, India}
		\author{ A. K. Pan }
	\email{akp@phy.iith.ac.in}
	\affiliation{National Institute of Technology Patna, Ashok Rajpath, Patna, Bihar 800005, India}
 \affiliation{Department of Physics, Indian Institute of Technology Hyderabad, Telengana-502284, India }

	\begin{abstract}
The nonlocality revealed in a multiparty multi-source network Bell experiment is conceptually different than the standard multiparty Bell nonlocality involving a single common source. Here, by introducing variants of asymmetric bilocal as well as trilocal network scenarios, we go beyond the typical bilocal network scenario where both the edge parties have an equal number of measurement settings. We first introduce an asymmetric bilocal network where one of the edge parties (say, Alice) receives $2^{n-1}$ inputs and the other edge party (say, Charlie) receives $n$ inputs. We derive two variants of asymmetric bilocality inequalities and demonstrate their optimal quantum violations. Further, we explore two types of asymmetric trilocal scenarios- (i) when two edge parties receive $2^{n-1}$ inputs each and the other edge party receives $n$ inputs, and (ii) when one edge party receives $2^{n-1}$ inputs and the other two edge parties have $n$ inputs each. We use an elegant sum-of-squares technique that enables us to evaluate the quantum optimal values of the proposed network inequalities without assuming the dimension of the systems for both the asymmetric bilocal as well as the trilocal scenarios. Further, we demonstrate the robustness of the quantum violations of the proposed inequalities in the presence of white noise. 
	\end{abstract}
	\maketitle


	\section{Introduction}\label{SecI}
	
	The study of quantum nonlocality in the network scenario \cite{Tavakoli2022} has recently been receiving considerable attention. Such a form of nonlocality is conceptually different from the standard Bell nonlocality \cite{Brunner2014}. While a multiparty Bell experiment involves a single common source, the multiparty network Bell experiment involves several independent sources. Each source distributes a physical system to subsequent parties, and each party performs a measurement on their subsystem prepared from different sources. 
	
	The simplest non-trivial network scenario \cite{Branciard2010, Fritz2012, Branciard2012} features three parties and two independent sources, commonly referred to as the bilocality scenario. The quantum nonlocality in a network is demonstrated through the quantum violation of suitably formulated nonlinear bilocality inequality. A straightforward generalization of the bilocality scenario is the $n$-locality scenario  \cite{Tavakoli2014, Andreoli2017, Tavakoli2016} involving an arbitrary $n$ number of sources. For example, a star network may have $n$ number of sources and edge parties. Each edge party shares the physical system with a central party. In recent times, the network nonlocality has been extensively studied in various topologies \cite{Rosset2016, Chaves2016, Tavakoli2016, Tavakoli2017, Gisin2017, Gisin2017Arxiv, Luo2018, Kerstjens2019, Lee2019, Krivachy2020, Supic2020PRL, Kundu2020, Roy2021, Tejada2021,   Jones2021, Renou2022}.
	
The reported interesting results such as possibility of observing quantum nonlocality without inputs \cite{Renou2019, Bancal2021} or showing the nonlocality of certain entangled states which do not exhibit nonlocality in the usual Bell scenario \cite{Sen(De)2005, Cavalcanti2011} establishes the fundamental importance of viewing nonlocality in terms of the symmetric network scenario in contrast to the standard Bell scenario. Characterization of network nonlocality and its correspondence with the bipartite Bell nonlocality has been studied \cite{Andreoli2017}. Several theoretical proposals have also been  experimentally verified \cite{Carvacho2017, Saunders2017, Andreoli2017, Sun2019, Poderini2020, baumer2021, Huang2022, Hakansson2022}. Recently, genuine network nonlocality has also been introduced that cannot be traced back to Bell nonlocality \cite{Tavakoli2021, Kerstjens2022, Renou2022}. Self-testing protocols using the quantum network have recently been proposed \cite{Agresti2021, Renou2021, Supic2022Arxiv, Supic2022}. Further, by using a quantum network, it has been established \cite{Renou2021, Li2022, Chen2022} that the real quantum theory can be experimentally falsified, i.e.,  quantum theory inevitably needs complex numbers. To this end, different forms of network scenarios like the star-network \cite{Tavakoli2014}, chain-shaped network \cite{Kundu2020} and cycle network \cite{Renou2019} have been explored.
	
We note here that while most of the studies concerning the star-network scenario have been investigated for the symmetric input scenario, i.e., each edge party performs the same number of measurements, a generalised study of network nonlocality in asymmetric input scenarios remains unexplored. In this regard, by introducing the asymmetric bilocal network scenario that comprises two edge parties performing three and six measurements respectively, and the central party performing four measurements, a couple of recent works \cite{Renou2021, Chen2022} have shown that complex numbers are necessary for quantum predictions. It is crucial to remark here that such a study is based on the two-qubit system. Here, the purpose of the present work is to probe hitherto unexplored generalised asymmetric network nonlocality in a device-independent way. In particular, by considering the bilocality scenario, we first derive asymmetric non-linear inequality for the scenario in which one of the two edge parties (say, Alice) has four measurement settings and the other edge party (say, Charlie) has three measurement settings. In addition to that, we extend the study from the asymmetric bilocality scenario to the asymmetric trilocality scenario involving three independent sources. In the trilocal network scenario, we explore two particular variants of asymmetric trilocality. First, we consider one edge party (Alice) has four measurement settings and the other two edge parties (Charlie and Diana) have three measurement settings each. Then we consider one edge party has three measurement settings (Charlie) and the other two edge parties have four measurement settings each (Alice and Diana).
	
	Then, by using an elegant sum-of-squares (SOS) approach developed in \cite{Pan2020, Munshi2021, Munshi2022}, we analytically obtain the optimal quantum violations of the asymmetric bilocality as well as the asymmetric trilocality inequalities. It is important to note that we evaluate such optimal quantum bounds without assuming the dimension of the systems. In this process, we evaluate the required constraints on the observables of each party for achieving the optimal quantum violation. For the asymmetric bilocality scenario, we demonstrate that the quantum optimal value will be achieved if each of the both Alice-Bob and Bob-Charlie shares at least a single copy of maximally entangled two-qubit state. Moreover, for the considered asymmetric trilocality scenario, as similar to the bilocality scenario, we find that the optimal quantum value will be obtained if each of all the edge parties shares at least a single copy of maximally entangled two-qubit state with the central party Bob. 
 
Furthermore, we extend our study for any arbitrary number of measurement settings. In particular, we consider an asymmetric bilocal network where one of the edge parties receives $2^{n-1}$ inputs and the other edge party receives $n$ inputs. We explore two types of asymmetric trilocal scenarios for arbitrary inputs.  (i) when two edge parties receive $2^{n-1}$ inputs each and the other edge party receives $n$ inputs, and (ii) when one edge party receives $2^{n-1}$ inputs and the other two edge parties have $n$ inputs each. Finally, we illustrate the robustness of the quantum violations of the proposed inequalities in the presence of the white noise for both the cases of bilocality and trilocality scenarios. We find that the proposed asymmetric inequality is most robust to white noise in the simplest bilocality network scenario.	
	
This paper is organized as follows. To begin with, in Sec. \ref{sosfors2}, by invoking the SOS approach \cite{Pan2020, Munshi2021}, we derive the optimal quantum violation of the standard bilocal network inequality without assuming the dimension of the system. Next, in Secs. \ref{secIII} and \ref{secIV}, we introduce two variants of asymmetric bilocal scenario and propose two different bilocal network inequalities. Then, we obtain the optimal quantum bounds along with the states and observables corresponding to the optimal quantum values (Secs. \ref{sos3s1} and \ref{sb3s1}). In  Secs. \ref{secV} and \ref{secVI}, by going beyond the bilocality network scenario, we introduce asymmetricity in the trilocal network and propose two different types of asymmetric trilocal inequalities. We also evaluate the corresponding optimal quantum bounds as well as the states and observables required for attaining such optimal quantum values (Secs. \ref{sost3} and \ref{st3s2}). In particular, we illustrate that in order to achieve the optimal quantum bound for both the asymmetric bilocal and trilocal cases, each of all the edge parties must share at least a single copy of maximally entangled state with the central party Bob. Then, in Sec.~\ref{secforn}, we have generalised the asymmetric bilocal and trilocal network scenario for arbitrary number of measurement settings. Further, in Sec.~(\ref{secVII}), we provide an analysis regarding the robustness of quantum violations of the proposed inequalities to white noise. Finally, in Sec. \ref{secVIII}, we discuss the salient features of our work and propose some interesting open questions.

	\section{Preliminaries: The standard bilocal network scenario} \label{secII}
	
The standard bilocal network scenario comprises of three spatially separated parties, Alice, Bob, and Charlie. Two independent sources $S_1$ and $S_2$ prepare a bipartite physical system for Alice-Bob and Bob-Charlie, respectively. Upon receiving the system from the respective source, Alice performs one of $m_A$ local measurements, denoted by $A_{n,x}\in \{A_{n,1},A_{n,2},..., A_{n,m_A}\}$.   Similarly for Bob and Charlie the respective measurements are denoted by $B_{n,j}\in \{B_{n,1},B_{n,2},..., B_{n,m_{B}}\}$ and  $C_{n,z}\in \{C_{n,1},C_{n,2},..., C_{n,m_C}\}$. The outcomes for Alice, Bob and Charlie are denoted by $a,b,c\in \{0,1\}$. The standard bilocal scenario is a symmetric scenario that implies an equal number of measurement settings for the edge parties, i.e., $m_A=m_C$. On the other hand, in our present work we consider two types of asymmetric bilocal network scenario - (i) Alice performs one of $m_A = 2^{n-1}$ measurements; Charlie and the central party Bob perform one of $m_C = m_B=n$ measurements. (ii) Alice and Bob perform one of $m_A =m_B=2^{n-1}$ measurements; Charlie performs one of $m_C=n$ measurements. The index `$n$' appearing in the subscript denotes the scenario involving the number of measurement settings considered. For example, $n=2$ corresponds the standard bilocal network scenario comprising two measurement settings for each party.
	
	Now, in the ontological model of the tripartite standard Bell scenario, it is assumed that the source prepares a common hidden variable $\lambda$. Then the reproducibility condition is given by   
	\begin{equation}
		\label{lcn}
		P(a,b,c|x,j,z)=\int d\lambda \ \mu (\lambda) P(a|x,\lambda) P(b|j,\lambda) P(c|z,\lambda) 
	\end{equation}

\begin{figure}[ht]
\centering
\includegraphics[width=1.0\linewidth]{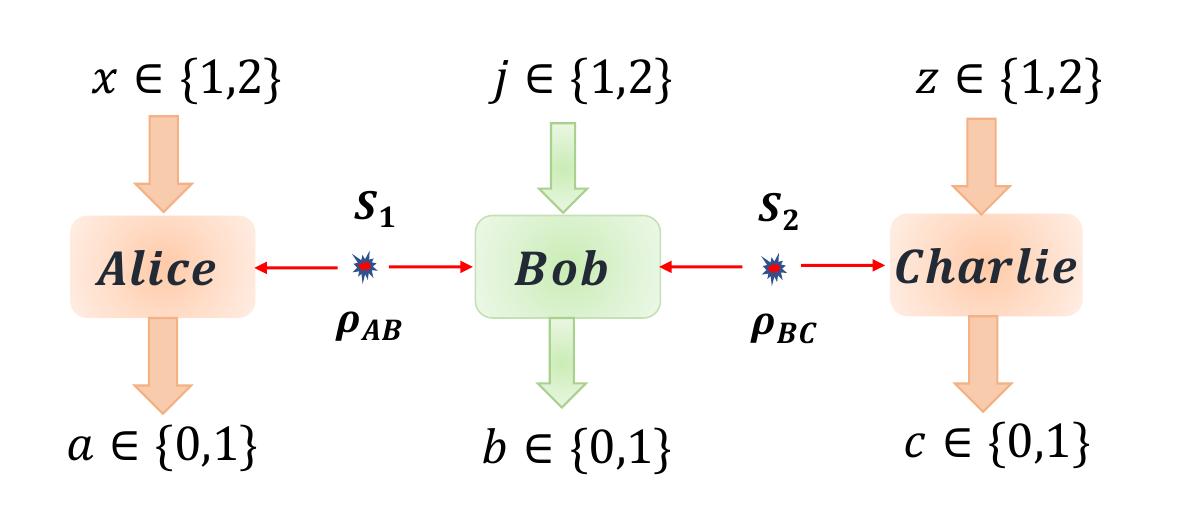}
\caption{The standard bilocal scenario featuring two edge parties (Alice and Charlie) and the central party Bob. The source $S_{1}(S_{2})$ emits physical system for Alice (Charlie) and Bob. The sources are assumed to be independent to each other.}
\label{bilocal}
\end{figure}
	
	In contrast to the ontological model of tripartite standard Bell scenario, the ontological model of bilocality scenario is that each source $S_{1}$ and $S_{2}$ prepares the physical system in the state $\lambda_{1}\in\Lambda_{1}$ and $\lambda_{2}\in\Lambda_{2}$ with a probability distribution $\mu (\lambda_{1})$ and $\mu(\lambda_{2})$ respectively, with $\int_{\Lambda_k}\mu (\lambda_{k})d\lambda_{k}=1$. The crucial assumption here is that the  sources $S_{1}$ and $S_{2}$ are  independent to each other. This means that  the joint probability distribution $\mu (\lambda_{1}, \lambda_{2})$ can be written in the factorized form as  $\mu (\lambda_{1}, \lambda_{2})= \mu (\lambda_{1})  \mu(\lambda_{2})$. Now, in order to reproduce the quantum theoretical prediction from the bilocal ontic model, the following reproducibility condition need to be satisfied.  

\begin{eqnarray}\label{blom}
P(a,b,c|x,j,z)&=&\iint d\lambda_{1} d\lambda_{2} \ \mu (\lambda_{1})  \mu(\lambda_{2})P(a|x,\lambda_{1}) 	\nonumber\\
&&  \hspace{1 cm} \times \ P(b|j,\lambda_{1},\lambda_{2}) P(c|z,\lambda_{2}) \nonumber\\
\end{eqnarray}

Note that in the 2 measurements per party scenario $(m_A=m_B=m_C=2)$, it has been shown \cite{Branciard2012} that any operational theory satisfying the above Eq.(\ref{blom}) satisfies the following nonlinear inequality
	\begin{equation}
		\label{bl22}
		\mathcal{B}_{2} \equiv \sqrt{|\mathcal{I}_{2,1}|}+ \sqrt{|\mathcal{I}_{2,2}|}\leq (\mathcal{B}_{2})_{bl}=2
	\end{equation}   
	where $\mathcal{I}_{2,2}$ are suitably defined linear combinations of the tripartite correlations, given by
	\begin{eqnarray}
		\label{ij22}
		\mathcal{I}_{2,1} &=&  \Big\langle \  \qty(A_{2,1}+ A_{2,2})\otimes B_{2,1} \otimes \qty(C_{2,1}+ C_{2,2}) \ \Big\rangle\label{s21}\\  
		\mathcal{I}_{2,2} &=&\Big\langle \ \qty(A_{2,1}- A_{2,2})\otimes B_{2,2} \otimes \qty(C_{2,2}-C_{2,1} ) \ \Big\rangle \label{s22}
	\end{eqnarray}
with\footnote{From now on we will denote  $\langle A_{n,x}\otimes B_{n,j}\otimes C_{n,z}\rangle$ as $\langle A_{n,x}B_{n,j}C_{n,z}\rangle$ }  $\langle A_{n,x}B_{n,j}C_{n,z}\rangle=\sum_{a,b,c} (-1)^{a+b+c}P(a,b,c|x,j,z)$ and $P(a,b,c|x,j,z)$ is the joint probability for obtaining the outcomes $(a,b,c)$ corresponding to the dichotomic measurements performed by Alice, Bob, and Charlie. In quantum theory, the joint probability $P(a,b,c|x,j,z)$ is given as
	\begin{equation}
		P(a,b,c|x,j,z)=\Tr[\qty(\rho_{AB}\otimes\rho_{BC}) \ \Pi_{A_{n,x}}^{a}\otimes\Pi_{B_{n,j}}^{b}\otimes\Pi_{C_{n,z}}^{c}]
	\end{equation}
	where $\rho_{AB}$ and $\rho_{BC}$ are bipartite states produced from two independent sources $S_1$ and $S_2$ respectively. 
	
	It has been shown \cite{Branciard2012, Munshi2021} that the maximum quantum value, $( \mathcal{B}_{2} )_{Q}^{opt}=2\sqrt{2}$ is obtained when Alice's and Charlie's observables are mutually anticommuting and Bob's observables are mutually commuting. An example of such choices of observables in two-dimensional Hilbert space $(\mathcal{H}^2)$ is given as follows
	\begin{eqnarray}\label{obs2}
		&&A_{2,1}=C_{2,1}= \left(\sigma_{z}+\sigma_{x}\right)/ \sqrt{2} \ ;\ \ B_{2,1}=\sigma_{z}\otimes\sigma_{z} \nonumber\\
		&& A_{2,2}=C_{2,2}=\left(\sigma_{z}-\sigma_{x}\right)/ \sqrt{2}  \ ; \ \ B_{2,2}=\sigma_{x}\otimes\sigma_{x} 
	\end{eqnarray}
	Note that while the optimal quantum value of $( \mathcal{B}_{2} )_{Q}^{opt}$ was earlier  derived \cite{Branciard2012} by taking a pair of the two-qubit entangled state, the dimension-independent derivation of $( \mathcal{B}_{2} )_{Q}^{opt}$ has recently been proposed \cite{Munshi2021}. Throughout this work, we adopt the SOS approach introduced in \cite{Pan2020} and derive the optimal quantum bound without assuming the dimension of the system. Thus, our optimal value possess the the potential to be used as device-independent certification of quantum correlations. 
	
	\subsection{Optimal quantum bound for $\mathcal{B}_{2}$} \label{sosfors2}
	
	Here, by invoking the elegant SOS approach \cite{Munshi2021}, we evaluate the optimal quantum  value of $(\mathcal{B}_{2})_{Q}$. Without loss of generality, one can always construct a suitable operator $\gamma_{2}$ satisfying $\langle\gamma_{2}\rangle = \beta_{2}- (\mathcal{B}_{2})_{Q} $ such that $\langle \gamma_2\rangle \geq0$. The existence of such operator $\gamma_2$ can be shown by suitably considering a set of operators $M_{2,j}, \ \forall j\in\{1,2\}$ which are polynomial functions of $A_{2,x}$, $B_{2,j}$ and $C_{2,z}$. 
	\begin{equation}
		\label{gamma2}
		\langle\gamma_{2}\rangle=\sum\limits_{j=1}^2 \frac{\sqrt{\omega_{2,j}}}{2}\left|M_{2,j}\ket{\psi} \right|^2
	\end{equation} 
	where $\omega_{2,j}\geq0$ is suitable positive numbers that  will be specified soon.  We choose the operator $M_{2,j}$ as the following way\footnote{From now on, we will write $A_{n,x}\otimes\mathbb{I}\otimes C_{n,z}$ as $A_{n,x}\otimes C_{n,z}$ and $\mathbb{I}_{d} \otimes B_{n,j} \otimes \mathbb{I}_{d}$ as $B_{n,j}$}
	\begin{eqnarray}
		&& \Big|M_{2,1}\ket{\psi} \Big|=\sqrt{\bigg|\left(\frac{{A}_{2,1}+{A}_{2,2}}{\omega_{2,1}^{A}}\otimes\mathbb{I}_{d}\otimes\frac{C_{2,1}+C_{2,2}}{\omega_{2,1}^{C}}\right)\ket{\psi}\bigg|} \nonumber\\
		&& \hspace{2cm} -\sqrt{\Big| \mathbb{I}_{d} \otimes B_{2,1} \otimes \mathbb{I}_{d} \ \ket{\psi}\Big|}\nonumber\\
		&&\Big|M_{2,2}\ket{\psi} \Big|=\sqrt{\bigg|\left(\frac{{A}_{2,1}-{A}_{2,2}}{\omega_{2,2}^{A}}\otimes\mathbb{I}_{d}\otimes\frac{C_{2,1}-C_{2,2}}{\omega_{2,2}^{C}}\right)\ket{\psi}\bigg|} \nonumber\\
		&& \hspace{2cm} -\sqrt{\Big| \mathbb{I}_{d} \otimes B_{2,2} \otimes \mathbb{I}_{d} \ \ket{\psi}\Big|}\label{mABC}
	\end{eqnarray}
	\begin{eqnarray}
		&&\omega_{2,1}^{A}=||({A}_{2,1}+{A}_{2,2})\ket{\psi}||_{2}=\sqrt{2+\langle\{A_{2,1},A_{2,2}\}\rangle} \nonumber \\
		&&\omega_{2,2}^{A}=||({A}_{2,1}-{A}_{2,2})\ket{\psi}||_{2}=\sqrt{2-\langle\{A_{2,1},A_{2,2}\}\rangle} \nonumber \\
		&&\omega_{2,1}^{C}=||({C}_{2,1}+{C}_{2,2})\ket{\psi}||_{2}=\sqrt{2+\langle\{C_{2,1},C_{2,2}\}\rangle} \nonumber\\
		&&\omega_{2,2}^{C}=||({C}_{2,1}-{C}_{2,2})\ket{\psi}||_{2}=\sqrt{2-\langle\{C_{2,1},C_{2,2}\}\rangle}\label{omega2j}
	\end{eqnarray}
Where $||  \cdot ||_2$ denotes the Frobenious norm given by $|| \ \mathcal{O} \ \ ||_2=\sqrt{\bra{\psi} \mathcal{O}^\dagger\mathcal{O}\ket{\psi}}$.	

	Now, since Alice, Bob and Charlie are space-like separated their observables are mutually commuting. Thus the operators ($A_{2,j}\otimes\mathbb{I}_d\otimes\mathbb{I}_d$), ($\mathbb{I}_d\otimes B_{2,j}\otimes\mathbb{I}_d$) and ($\mathbb{I}_d\otimes\mathbb{I}_d\otimes C_{2,j}$) are also mutually commuting. Hence, these three observables must have at least one common eigenstate. Without loss of generality, $\ket{\psi}$ is taken to be one of the common eigenstate. Therefore, by evaluating the quantity $|M_{2,j}\ket{\psi}|^2$ from Eq. (\ref{mABC}), a simple algebraic manipulation gives us from Eq. (\ref{gamma2}) the following
	\begin{eqnarray} \label{sos12}
		\langle\gamma_{2}\rangle &=& \qty(\sqrt{\omega_{2,1}}+\sqrt{\omega_{2,2}})- (\mathcal{B}_2)_Q
	\end{eqnarray} 
	where $\omega_{2,j}= \omega_{2,j}^A \ \omega_{2,j}^C$.
	
	Now, since by construction $\langle\gamma_{2}\rangle \geq0$, it is evident that from the above Eq. (\ref{sos12}) that the optimal quantum value of $\mathcal{B}_{2}$ corresponds to $\langle\gamma_{2}\rangle =0$. Therefore, the quantum optimal value is given as follows 
	\begin{equation}\label{SQopt}
		(\mathcal{B}_{2})_{Q}^{opt}=\sqrt{\omega_{2,1}^A \ \omega_{2,1}^C}+\sqrt{\omega_{2,2}^A \ \omega_{2,2}^C}
	\end{equation}
	
	Next, using the inequality, 
	$\sqrt{r_{1}s_{1}}+\sqrt{r_{2}s_{2}}\leq\sqrt{r_{1}+r_{2}}\sqrt{s_{1}+s_{2}} \ \ \forall r_{1}, s_{1}, r_{2}, s_{2}\geq 0$, we can write Eq.(\ref{SQopt}) as
	\begin{eqnarray}\label{dd}		
		(\mathcal{B}_{2})_{Q}^{opt}&=&  \sqrt{\left(\omega_{2,1}^{A}+\omega_{2,2}^{A}\right)\left(\omega_{2,1}^{C}+\omega_{2,2}^{C}\right)} \nonumber \\
		&=& \sqrt{\qty(\sqrt{2+\langle \{A_{2,1},A_{2,2}\}\rangle}+\sqrt{2-\langle \{A_{2,1},A_{2,2}\}\rangle})} \nonumber\\
		&\times& \sqrt{\qty(\sqrt{2+\langle \{C_{2,1},C_{2,2}\}\rangle}+\sqrt{2-\langle \{C_{2,1},C_{2,2}\}\rangle})}  \nonumber\\
	\end{eqnarray}
	
	The above Eq.~(\ref{dd}) is optimised when $\{A_{2,1},A_{2,2}\}=\{C_{2,1},C_{2,2}\}=0$ and the optimal quantum value is then given by
	\begin{equation}
		(\mathcal{B}_2)_Q^{opt} = 2\sqrt{2}
	\end{equation}
	
	Note that the evaluated optimal quantum value $( \mathcal{B}_{2} )_{Q}^{opt}=2\sqrt{2}$ is same as that obtained when the state between three parties are assumed to be pair of maximally entangled two-qubit state. This means that for the 2-settings bilocality scenario, the optimal quantum violation of the bilocal inequality remain same even if one consider higher dimensional maximally entangled states.


\section{Asymmetric bilocal network: scenario-I}\label{secIII}
	
\begin{figure}[ht]
\centering
\includegraphics[width=0.8\linewidth]{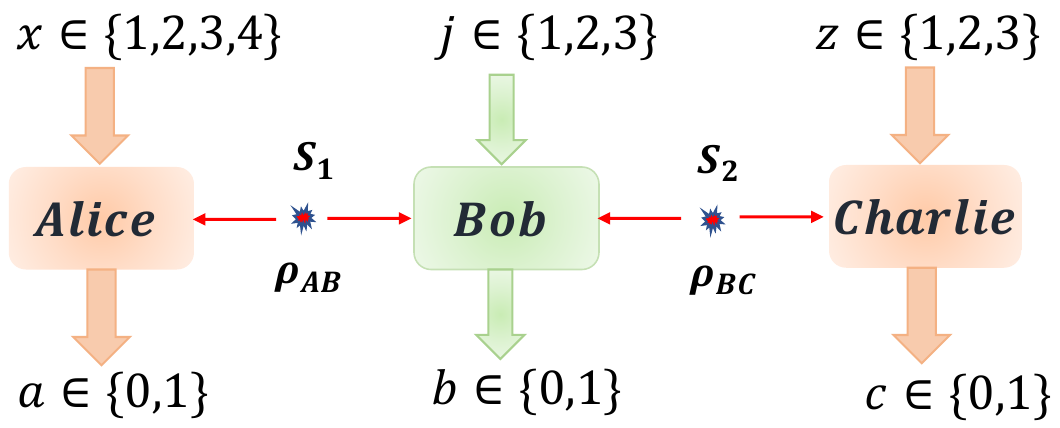}
\caption{Asymmetric bilocal network scenario featuring two edge parties (Alice and Charlie) and a central party Bob. The independent sources $S_{1}$ and $S_{2}$ emit physical systems for Alice-Bob and Charlie-Bob respectively.}
\label{b2s1}
\end{figure}
	
Here we introduce a variant of asymmetric bilocal network scenario for $n=3$, in which Alice performs one of the four dichotomic measurements and the other edge party (Charlie) perform one of three dichotomic measurements. For our purpose, we consider that the central party performs one of three dichotomic measurements. Note that for the 2-settings bilocality scenario, is same as the symmetric bilocal scenario, where each party has equal number (two) of measurement settings. In the following, we derive a family of non-linear asymmetric bilocality inequalities and also evaluate their optimal quantum violations.
	
	In this scenario, let us consider the nonlinear bilocal inequality of the following form
	\begin{equation}
		\label{n3}
		\mathcal{B}_{3}=\sum_{j=1}^{3}\sqrt{\qty|\mathcal{I}_{3,j}|} \ \leq \ \left(\mathcal{B}_{3}\right)_{bl}
	\end{equation}
	where $\left(\mathcal{B}_{3}\right)_{bl}$ is the bilocal bound of $\mathcal{B}_{3}$ and the quantity $\mathcal{I}_{3,j}=\Big\langle \mathcal{\tilde{A}}_{3,j}B_{3,j}\mathcal{\tilde{C}}_{3,j}\Big\rangle$. We define the quantities $\mathcal{\tilde{A}}_{3,j}$ and $\mathcal{\tilde{C}}_{3,j}$ as
	\begin{eqnarray}\label{Inj}
		\mathcal{\tilde{A}}_{3,1}&=&A_{3,1}+A_{3,2}+ A_{3,3}-A_{3,4} \nonumber \\
		\mathcal{\tilde{A}}_{3,2}&=&A_{3,1}+A_{3,2}-A_{3,3}+A_{3,4} \nonumber\\
		\mathcal{\tilde{A}}_{3,3}&=&A_{3,1}-A_{3,2}+A_{3,3}+A_{3,4}  \nonumber \\
		\mathcal{\tilde{C}}_{3,j}&=& C_{3,j}+C_{3,j+1} \ \ \text{with} \  C_{3,4}=-C_{3,1}
	\end{eqnarray}
	
	In an ontological model $\lambda_1 \in \Lambda_1$ and $\lambda_2 \in \Lambda_2$ completely determines the statistics of all the measurements. This means that in the ontological model, we can write the following
	\begin{eqnarray} \label{olm3}
		\langle A_{3,x}\rangle_{\lambda_{1}} &=& \sum_{a}(-1)^{a}P(a|x,\lambda_{1}) \ \ \ \ \forall x\in\{1,2,3,4\}\nonumber \\
		\langle C_{3,z}\rangle_{\lambda_{2}} &=& \sum_{c}(-1)^{c}P(c|z,\lambda_{2})  \ \ \ \ \forall z\in\{1,2,3\}\nonumber \\
		\langle B_{3,j}\rangle_{\lambda_{1},\lambda_{2}} &=& \sum_{b}(-1)^{b}P(b|j,\lambda_{1},\lambda_{2})   \ \ \forall j\in\{1,2,3\}
	\end{eqnarray}
	
	Now, invoking the reproducibility condition given by Eq.~(\ref{blom}) it follows that
	\begin{eqnarray}
		\label{bilocal31}
		\mathcal{I}_{3,1}&=&\iint d\lambda_{1} d\lambda_{2} \ \mu (\lambda_{1}) \mu (\lambda_{2}) \ \qty[ \langle C_{3,1}\rangle_{\lambda_{2}}+\langle C_{3,2}\rangle_{\lambda_{2}} ] \ \langle B_{3,1}\rangle_{\lambda_{1},\lambda_{2}} \nonumber\\
		&& \   \times \ \ \qty[ \langle A_{3,1}\rangle_{\lambda_{1}}+\langle A_{3,2}\rangle_{\lambda_{1}}+\langle A_{3,3}\rangle_{\lambda_{1}}-\langle A_{3,4}\rangle_{\lambda_{1}}] 
	\end{eqnarray} 
	Since $|\langle B_{3,1}\rangle_{\lambda_{1},\lambda_{2}}|\leq1$, we obtain
	\begin{eqnarray}\label{3cond}
		\qty|\mathcal{I}_{3,1}|&\leq& \iint d\lambda_{1} d\lambda_{2} \ \mu (\lambda_{1}) \mu (\lambda_{2}) \  \qty| \langle C_{3,1}\rangle_{\lambda_{2}}+\langle C_{3,2}\rangle_{\lambda_{2}}| \nonumber\\
		&& \  \times \ \ \qty| \langle A_{3,1}\rangle_{\lambda_{1}}+\langle A_{3,2}\rangle_{\lambda_{1}}+\langle A_{3,3}\rangle_{\lambda_{1}}-\langle A_{3,4}\rangle_{\lambda_{1}}| 
	\end{eqnarray}
	
	The terms $|\mathcal{I}_{3,2}|$ and $|\mathcal{I}_{3,3}|$ can also be written in a similar manner as Eq. (\ref{3cond}).  Now for our purpose, by utilizing the inequality\footnote{\label{f2} The inequality proved in Appendix A of \cite{Tavakoli2014} 
		\begin{equation}
			\sum\limits_{i=1}^{t}\bigg(\prod\limits_{k=1}^{r}z_{k}^{i}\bigg)^{\frac{1}{r}}\leq \prod \limits_{k=1}^{r}\bigg(\sum\limits_{i=1}^{t}z_{k}^{i}\bigg)^{\frac{1}{r}} \ \ \ \forall z_{k}^{i} \geq 0 \nonumber 
		\end{equation} Here r is the number of edge party in a star-network. Note that for our bilocal case $r=2$.} proved in \cite{Tavakoli2014}, we obtain the following	
	\begin{equation}\label{S3}
		(\mathcal{B}_{3})_{bl}\leq \qty(\iint d\lambda_{1} d\lambda_{2} \ \mu(\lambda_{1})\mu(\lambda_{2}) \ \delta_{1} \ \delta_{2})^{\frac{1}{2}}
	\end{equation}
	where  $\delta_{1}=\big[|\langle{A_{3,1}}\rangle_{\lambda_{1}}+\langle{A_{3,2}}\rangle_{\lambda_{1}}+\langle{A_{3,3}}\rangle_{\lambda_{1}}- \langle{A_{3,4}}\rangle_{\lambda_{1}}|+|\langle{A_{3,1}}\rangle_{\lambda_{1}}+\langle{A_{3,2}}\rangle_{\lambda_{1}}-\langle{A_{3,3}}\rangle_{\lambda_{1}}+ \langle{A_{3,4}}\rangle_{\lambda_{1}}|+|\langle{A_{3,1}}\rangle_{\lambda_{1}}-\langle{A_{3,2}}\rangle_{\lambda_{1}}+\langle{A_{3,3}}\rangle_{\lambda_{1}}+\langle{A_{3,4}}\rangle_{\lambda_{1}}|\big]$ and $\delta_{2}=\big[|\langle{C_{3,1}}\rangle_{\lambda_{2}}+\langle{C_{3,2}}\rangle_{\lambda_{2}}|+|\langle{C_{3,2}}\rangle_{\lambda_{2}}+\langle{C_{3,3}}\rangle_{\lambda_{2}}|+|\langle{C_{3,3}}\rangle_{\lambda_{2}}-\langle{C_{3,1}}\rangle_{\lambda_{2}}| \big]$. 
	
	Since all the observables are dichotomic with eigenvalues $\pm1$, it is straightforward to derive that $\delta_{1}\leq6$ and $\delta_{2}\leq 4$. Therefore, from Eq.~(\ref{S3}), integrating over $\lambda_{1}$ and $\lambda_{2}$ we obtain 
	\begin{equation}\label{S3fc}
		(\mathcal{B}_{3})_{bl}\leq 2\sqrt{6}\approx 4.89
	\end{equation}
We show that there are suitable states and observables for which the bilocal bound can be violated in quantum theory. Now, in the following, we evaluate the quantum optimal bound of $\mathcal{B}_{3}$.

\subsection{Optimal quantum bound of the asymmetric bilocality inequality in scenario-I }\label{sos3s1}

To derive the optimal quantum bound of $\mathcal{B}_{3}$ without assuming the dimension of the system,  we again invoke the SOS approach discussed in the preceding Sec.~\ref{sosfors2}. Following the similar argument presented earlier in the Sec.~\ref{sosfors2}, we first show that there exists a suitable operator $\gamma_{3}$ satisfying $\langle\gamma_{3}\rangle = \beta_{3}- (\mathcal{B}_{3})_{Q} $ such that $\langle \gamma_3\rangle \geq0$. Note that the power of SOS approach in evaluating the  optimal quantum bound of a particular Bell functional lies in the suitable construction of the operator $\gamma_{3}$ in such a way so that it is a semi-definite operator and can be reduced to the form of the concerned Bell functional. Now, in order to evaluate the quantum optimal bound of $\mathcal{B}_3$, we construct $\gamma_3$ in terms of a set of operators $M_{3,j} \ \forall j\in \{1,2,3\}$ in the following way
	\begin{equation}
		\label{gamma3j}
		\langle\gamma_{3}\rangle=\sum\limits_{j=1}^3 \frac{\sqrt{\omega_{3,j}}}{2}\qty|M_{3,j}\ket{\psi}|^2
	\end{equation} 
	where $\omega_{3,j}\geq0$ is suitable positive numbers and $\omega_{3,j}=\omega^{A}_{3,j}\cdot\omega^{C}_{3,j}$ that  will be specified soon.  In the following $\forall j\in\{1,2,3\}$, we choose the operator $M_{3,j}$ and the quantity $\omega_{3,j}$.
	\begin{eqnarray} 
		&&\qty|M_{3,j}\ket{\psi}|=\sqrt{\qty| \qty(\frac{\mathcal{\tilde{A}}_{3,j}}{\omega_{3,j}^{A}}\otimes\frac{\mathcal{\tilde{C}}_{3,j}}{\omega_{3,j}^{C}})\ket{\psi}|} -\sqrt{\qty|B_{3,j}\ket{\psi}|} \label{m31} \\
		&& \omega_{3,j}^A=||\mathcal{\tilde{A}}_{3,j}\ket{\psi}||_2 \ ; \ \ \omega_{3,j}^C=||\mathcal{\tilde{C}}_{3,j}\ket{\psi}||_2 \label{omega31s1}
	\end{eqnarray}	
	
Putting $M_{3,j}$ from Eq.~(\ref{m31}) into the Eq. (\ref{gamma3j}), after a simple algebraic evaluation, we obtain $ \langle \gamma_3 \rangle =\sum\limits_{j=1}^3 \sqrt{\omega_{3,j}}-(\mathcal{B}_3)_Q$. Then, it follows that the quantum optimal value corresponds $\langle \gamma_3 \rangle=0$. Therefore,
	\begin{eqnarray}\label{SQopt3}
		\left(\mathcal{B}_{3}\right)_{Q}^{opt} &=& \max \sum\limits_{j=1}^3 \sqrt{\omega_{3,j}^{A}\cdot\omega_{3,j}^{C}}
	\end{eqnarray}
which in turn gives the optimisation condition as follows
	\begin{equation}\label{impn3s1}
		\qty|M_{3,j}\ket{\psi}|=0 \ \implies M_{3,j}\ket{\psi}=0 \ \ \forall j \in \{1,2,3\}
	\end{equation}
	
	Now, in order to evaluate the optimal quantum value $\left(\mathcal{B}_{3}\right)_{Q}^{opt}$ and thus the quantity $\sum\limits_{j=1}^{3}\sqrt{\omega_{3,j}^A \cdot \omega_{3,j}^C}$ \ , we invoke the inequality given in the footnote \ref{f2}. Then the right-hand-side of Eq. (\ref{SQopt3}) reduces to the following 
	\begin{equation}
		\label{omega1s1}
		\sum\limits_{j=1}^{3}\left(\prod \limits_{k=A,C}\omega^{k}_{3,j}\right)^{\frac{1}{2}}\leq\prod \limits_{k=A,C}\left(\sum\limits_{j=1}^{3}\omega^{k}_{3,j}\right)^{\frac{1}{2}}	
	\end{equation}
	
	Further, by applying the convex inequality\footnote{\label{f3} From the Jensen's inequality given by $f(\sum\limits_{k=1}^{t}r_k x_k) \leq \sum\limits_{k=1}^{t} r_k f(x_k) $ where $\sum\limits_{k=1}^{t}r_k =1$, the following inequality can be derived \begin{equation} \sum\limits_{k=1}^{t}\omega_{k}\leq \sqrt{\ t\sum\limits_{k=1}^{t}\omega^2_k}\nonumber \end{equation}}, the quantity $\sum\limits_{j=1}^{3}\omega^{k}_{3,j}$  can be written as
	\begin{equation}
		\label{omega2s1}
		\sum\limits_{j=1}^{3}(\omega^{k}_{3,j})\leq \sqrt{3\sum\limits_{j=1}^{3}\bigg(\omega^k_{3,j}\bigg)^{2}}
	\end{equation}
	
	Then, by combining Eq. (\ref{omega1s1}) and (\ref{omega2s1}), from Eq. (\ref{SQopt3} ) we obtain 
	\begin{eqnarray}\label{s3m1}
		\left(\mathcal{B}_{3}\right)_{Q}^{opt} = \max\left[\prod_{k=A,C}^{}\left(3\sum_{j=1}^{3}\left( \omega^{k}_{3,j}\right)^2\right) \right] ^{\frac{1}{4}}
	\end{eqnarray}
	
	where each $\left( \omega^{k}_{3,j}\right)^2$ is evaluated from Eq. (\ref{omega31s1}) as follwos	\begin{eqnarray} 
		\left( \omega^{A}_{3,1}\right)^2&=&\bra{\psi}(4+\{ A_{3,1},(A_{3,2}+A_{3,3}-A_{3,4}) \}\nonumber\\
		&+&\{A_{3,2},(A_{3,3}-A_{3,4})\}-\{A_{3,3},A_{3,4}\})\ket{\psi} \label{w31}\nonumber\\
		\left( \omega^{A}_{3,2}\right)^2&=&\bra{\psi}(4+\{A_{3,1},(A_{3,2}-A_{3,3}+A_{3,4}) \}\nonumber\\
		&+&\{A_{3,2},(-A_{3,3}+A_{3,4})\}-\{A_{3,3},A_{3,4} \}) \ket{\psi}\label{w32}\nonumber\\
		\left( \omega^{A}_{3,3}\right)^2&=&\bra{\psi}(4+\{A_{3,1},(-A_{3,2}+A_{3,3}+A_{3,4}) \}\nonumber\\
		&-&\{A_{3,2},(A_{3,3}+A_{3,4})\}+\{A_{3,3},A_{3,4} \})\ket{\psi}\label{w33}\\
		\left( \omega^{C}_{3,1}\right)^2&=&\bra{\psi}(2+\{C_{3,1},C_{3,2}\})\ket{\psi}\label{w311}\nonumber\\
		\left( \omega^{C}_{3,2}\right)^2&=&\bra{\psi}(2+\{C_{3,2},C_{3,3}\})\ket{\psi}\label{w322}\nonumber\\
		\left( \omega^{C}_{3,3}\right)^2&=&\bra{\psi}(2-\{C_{3,1},C_{3,3}\})\ket{\psi}\label{w333}
	\end{eqnarray}
	
	Now, in the following we calculate $\sum_{j=1}^{3}\left( \omega^{A}_{3,j}\right)^2$ from Eq. (\ref{w33})  and  $\sum_{j=1}^{3}\left( \omega^{C}_{3,j}\right)^2$ from Eq. (\ref{w333}) separately.\\
	
	\paragraph{\textit{Evaluation of $\sum_{j=1}^{3}\left( \omega^{A}_{3,j}\right)^2$:}}\label{pws1}
	
	\begin{eqnarray}\label{v31}
		\sum_{j=1}^{3}\left( \omega^{A}_{3,j}\right)^2&=& \bra{\psi}(12+ \{A_{3,1},(A_{3,2}-A_{3,3}+A_{3,4}) \} \nonumber\\
		&-&\{A_{3,2},(A_{3,3}+A_{3,4})\} - \{A_{3,3},A_{3,4} \})\ket{\psi}\nonumber\\
		&=&\bra{\psi}(12+  \Delta_{3})\ket{\psi}
	\end{eqnarray} 
	where $\Delta_{3}=\{A_{3,1},(A_{3,2}-A_{3,3}+A_{3,4}) \} -\{A_{3,2},(A_{3,3}+A_{3,4})\}-\{A_{3,3},A_{3,4} \}$.
	Without loss of generality we can always write $\ket{\psi^{\prime}}=(A_{3,1}+A_{3,2}+A_{3,3}-A_{3,4}) \ket{\psi}$ such that $\ket{\psi}\neq0$. Therefore, $\braket{\psi^{\prime}|\psi^{\prime}}=\bra{\psi}(4-\Delta_{3})\ket{\psi}$ implies $\braket{\Delta_{3}} =4- \braket{\psi^{\prime}|\psi^{\prime}} $. Then it immediately follows that $\braket{\Delta_{3}}_{max}$ is obtained \textit{iff} $\braket{\psi^{\prime}|\psi^{\prime}}=0$. Since $\ket{\psi}\neq0$, then the following relation must satisfy
	\begin{equation}
		\label{condAlices1}
		A_{3,1}-A_{3,2}-A_{3,3}-A_{3,4}=0
	\end{equation}
	Hence, in order to obtain the quantum optimal value of $(\mathcal{B}_{3})^{opt}_Q$, observables of Alice must satisfy the linear condition given by Eq. (\ref{condAlices1}). Therefore, $\langle\Delta_{3}\rangle_{max}=4$ leads to   
	\begin{equation}
		\label{omega3as1}
		\sum_{j=1}^{3}\left( \omega^{A}_{3,j}\right)^2=\bra{\psi}(12+  \Delta_{3})\ket{\psi}\leq 16	
	\end{equation} 
	
	\paragraph{\textit{Evaluation of $\sum_{j=1}^{3}\left( \omega^{C}_{3,j}\right)^2$:}}\label{pws1c}

	\begin{eqnarray}\label{v32}
		\sum_{j=1}^{3}\left( \omega^{C}_{3,j}\right)^2&=& \bra{\psi}(6+ \{C_{3,2},(C_{3,1}+C_{3,3})\}-\{C_{3,1},C_{3,3} \})\ket{\psi}\nonumber\\
		&=& \bra{\psi}(6+ 3\mathbb{I}-(C_{3,1}-C_{3,2}+C_{3,3})^2)\ket{\psi} \leq 9 
	\end{eqnarray}
	The above Eq. (\ref{v32}) is maximized when 
	\begin{equation} \label{v32conds1}
		C_{3,1}-C_{3,2}+C_{3,3}=0
	\end{equation}
	Finally, we obtain the optimal quantum value from Eqs. (\ref{SQopt3}), (\ref{omega3as1}) and (\ref{v32}) as 
	\begin{eqnarray}\label{s3qm}
		\left(\mathcal{B}_{3}\right)_{Q}^{opt}=6	
	\end{eqnarray}
	
	It is important to remark here that the optimal quantum value $\left(\mathcal{B}_{3}\right)_{Q}^{opt}=6$ is evaluated without specifying the dimension of both the system and observables. The optimal value fixes the states, and the observables are the following. 
	
	\subsection{The state and observables for the optimal quantum violation of $\mathcal{B}_{3}$}\label{sb3s1}
	
	We further obtain relationships between the observables of all the parties for achieving the optimal quantum violation. Such relationships are given in Eqs.~(\ref{condAlices1}) and Eq.~(\ref{v32conds1}), which in turn provide the relationship between the observables in terms of the anti-commuting relations.
	
	\begin{eqnarray}
		&&\{A_{3,1},A_{3,2}\}=\{A_{3,1},A_{3,3}\}=\{A_{3,1},A_{3,4}\}=\frac{2}{3} \mathbb{I}_{d}\nonumber\\
		&&\{A_{3,2},A_{3,3}\}=\{A_{3,2},A_{3,4}\}=\{A_{3,3},A_{3,4}\}=-\frac{2}{3} \mathbb{I}_{d} \nonumber\\
		&&\{C_{3,1},C_{3,2} \}=\{C_{3,2},C_{3,3} \}=-\{C_{3,1},C_{3,3}\}= \mathbb{I}_{d} \label{charliecond4}
	\end{eqnarray}
	
	By using the above relations between the observables  given by Eq.~(\ref{charliecond4}) on the observables, one can always construct a set of observables for Alice and Charlie in the Hilbert space dimension $\mathcal{H}^d , \ \ \forall d\geq2$.
	
	Next, we recall the optimization condition obtained in the SOS method from the Eq.~(\ref{impn3s1}) to find the constraints on Bob's observable. The specific condition $M_{3,j} \ket{\psi} =0 \ ; \ \forall j\in\{1,2,3\}$ implies the following
	\begin{eqnarray} \label{bobcond3}
		B_{3,j}&=&\frac{\mathcal{\tilde{A}}_{3,j}}{\omega_{3,j}^{A}}\otimes\frac{\mathcal{\tilde{C}}_{3,j}}{\omega_{3,j}^{C}}
	\end{eqnarray}
	
	In the following, we then explicitly construct a set of observables for the Hilbert space dimension $\mathcal{H}^2$
	\begin{eqnarray}\label{obsn3}
		C_{3,1}&=&\sigma_{z} \ ; \  C_{3,2}=\left(\frac{\sqrt{3}}{2}\sigma_{x}+\frac{\sigma_{z}}{2}\right) ; \  C_{3,3}=\left(\frac{\sqrt{3}}{2}\sigma_{x}-\frac{\sigma_{z}}{2}\right) \nonumber \\
		A_{3,1}&=& \frac{\sigma_{x}+\sigma_{y}+\sigma_{z}}{\sqrt{3}} \ \ ;\ \ \ \ A_{3,2}=\frac{\sigma_{x}+\sigma_{y}-\sigma_{z}}{\sqrt{3}} \nonumber\\
		A_{3,3}&=& \frac{\sigma_{x}-\sigma_{y}+\sigma_{z}}{\sqrt{3}} \ \ ;\ \ \ \ A_{3,4}=\frac{-\sigma_{x}+\sigma_{y}+\sigma_{z}}{\sqrt{3}}
	\end{eqnarray}
	Note that employing the above-mentioned observables, we find that the quantum optimal value $\left(\mathcal{B}_{3}\right)_{Q}^{opt}=6$ is achieved when two maximally entangled two-qubit states are shared between Alice-Bob and Bob-Charlie.\\


	\section{Asymmetric bilocal network: scenario-II}\label{secIV}
	
	\begin{figure}[ht]
		\centering
		\includegraphics[width=0.8\linewidth]{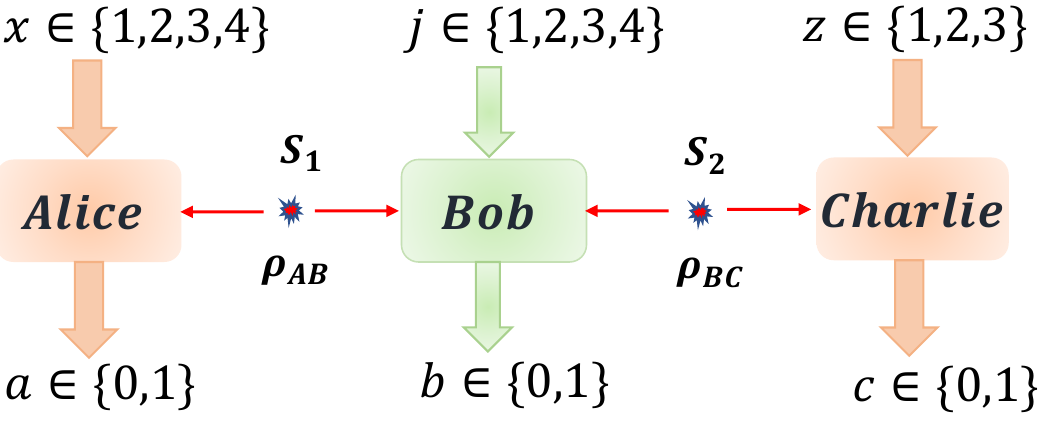}
		\caption{Asymmetric bilocal network scenario featuring two edge parties (Alice and Charlie) and a central party Bob. The independent sources $S_{1}$ and $S_{2}$ emit physical systems for Alice-Bob and Charlie-Bob respectively.}
		\label{b2s2}
	\end{figure}
	
	Here we consider the central party Bob performs equal number of measurements (four) as Alice in contrast to the preceding scenario discussed in Sec. \ref{secIII} where number of measurements  for Bob and Charlie were considered to be equal. In this scenario, let us consider the nonlinear bilocal inequality of the following form
	\begin{equation}
		\label{n31s2}
		\mathcal{B}_{3}^{\prime}=\sum_{j=1}^{4}\sqrt{ \ \qty| \ \Big\langle \mathcal{\tilde{A}}_{3,j}^{\prime}B_{3,j}\mathcal{\tilde{C}}_{3,j}^{\prime}\Big\rangle \ | \ }\leq (\mathcal{B}_{3}^{\prime})_{bl}
	\end{equation}	
	where $(\mathcal{B}_{3}^{\prime})_{bl}$ is the bilocal bound of $\mathcal{B}_{3}^{\prime}$ and $\mathcal{\tilde{A}}_{3,j}^{\prime} = A_{3,j}+A_{3,j+1}$ satisfying $A_{3,4}=-A_{3,1}$; $\mathcal{\tilde{C}}_{3,1}^{\prime} =C_{3,1}+C_{3,2}+C_{3,3}$; $\mathcal{\tilde{C}}_{3,2}^{\prime} =C_{3,1}+C_{3,2}-C_{3,3}$; $\mathcal{\tilde{C}}_{3,3}^{\prime}=C_{3,1}-C_{3,2}+C_{3,3}$;$\mathcal{\tilde{C}}_{3,4}^{\prime} = -C_{3,1}+C_{3,2}+C_{3,3}$.
	
	We find (see appendix \ref{aps21}) that the bilocal bound in this scenario is  $(\mathcal{B}_{3}^{\prime})_{bl}=6$.
	It can be shown that there are suitable states and observables for which quantum correlations violate the bilocal bound. In the following, we evaluate the quantum optimal bound of $\mathcal{B}_{3}^{\prime}$.
	
	
	The optimal quantum bound of $\mathcal{B}_{3}^{\prime}$ without assuming the dimension of the system is derived by suitably invoking the SOS approach. The explicit construction of the SOS method is provided in Appendix (\ref{aps22}). The optimal quantum bound is found to be 
	\begin{eqnarray}
		\qty(\mathcal{\mathcal{B}}_{3}^{\prime})_{Q}^{opt}=4\left[3\left(2+\sqrt{2}\right)\right]^{\frac{1}{4}} \approx 7.16
	\end{eqnarray} 
	
	It is to be noted here that the obtained optimal quantum bound $\qty((\mathcal{\mathcal{B}}_{3}^{\prime})_{Q}^{opt}\approx 7.16)$ in this scenario is greater than that obtained $\qty((\mathcal{\mathcal{B}}_{3})_{Q}^{opt}=6)$ in the earlier scenario-I. Thus, the bilocal as well as the quantum optimal bound depends on how the asymmetry invoked in the bilocal scenario. The corresponding state and observables for which the quantum optimal bound is achieved is given in Appendix (\ref{aps23}).
	
	
	\section{Asymmetric trilocal network: scenario-I}\label{secV}
	
	\begin{figure}[ht]
		\centering
		\includegraphics[width=0.6\linewidth]{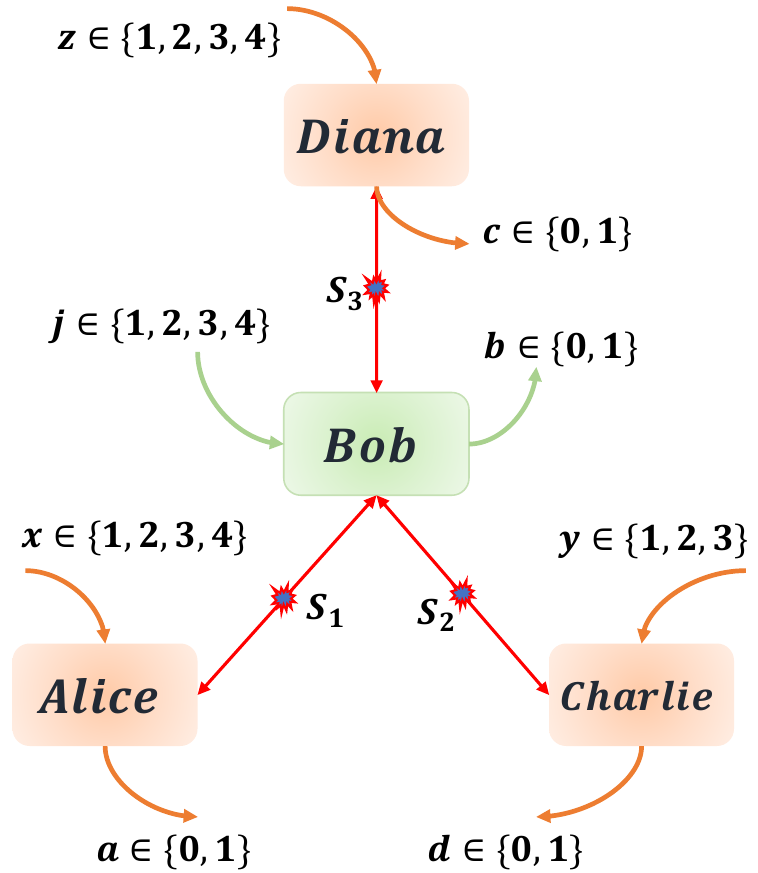}
		\caption{Asymmetric trilocal network scenario featuring three edge parties (Alice, Charlie, and Diana) and a central party Bob. The independent sources $S_{1}$, $S_{2}$ and $S_{3}$ emit physical systems for Alice-Bob, Charlie-Bob, and Diana-Bob respectively.}
		\label{t1s1}
	\end{figure}

	Here we introduce a variant of asymmetric trilocal network scenario in which one of the edge party Charlie performs one of three dichotomic measurements and rest of the edge parties Alice and Diana as well as the central party Bob perform four dichotomic measurements. In this scenario, let us introduce the nonlinear trilocal inequality of the following form
	\begin{equation}
		\label{t3}
		\mathcal{T}_{3}=\sum_{j=1}^{4}|\mathcal{J}_{3,j}|^{\frac{1}{3}}\leq (\mathcal{T}_{3})_{tl}
	\end{equation}	
	where $\left(\mathcal{T}_{3}\right)_{tl}$ is the local bound of $\mathcal{T}_{3}$ and the quantity $\mathcal{J}_{3,j}=\Big\langle \mathscr{\tilde{A}}_{3,j}B_{3,j}\mathscr{\tilde{C}}_{3,j} \mathscr{\tilde{D}}_{3,j}\Big\rangle$. We define the quantities $\mathscr{\tilde{A}}_{3,j}$, $\mathscr{\tilde{C}}_{3,j}$ and $\mathscr{\tilde{D}}_{3,j}$ as follows
	\begin{eqnarray}\label{j31}
		&&	\mathscr{\tilde{A}}_{3,j} = A_{3,j}+A_{3,j+1}  \ \ \text{with} \ A_{3,5}=-A_{3,1} \nonumber \\
		&&	\mathscr{\tilde{C}}_{3,1} =C_{3,1}+C_{3,2}+C_{3,3} \ ; \ \	\mathscr{\tilde{C}}_{3,2} =C_{3,1}+C_{3,2}-C_{3,3} \nonumber \\	
		&&	\mathscr{\tilde{C}}_{3,3}=C_{3,1}-C_{3,2}+C_{3,3}  \ ; \ \
		\mathscr{\tilde{C}}_{3,4}= -C_{3,1}+C_{3,2}+C_{3,3} \nonumber\\
		&&\mathscr{\tilde{D}}_{3,1}=D_{3,1}+D_{3,2}+D_{3,3}+D_{3,4}\nonumber\\
		&&\mathscr{\tilde{D}}_{3,2}=D_{3,1}+D_{3,2}+D_{3,3}-D_{3,4}\nonumber\\
		&&\mathscr{\tilde{D}}_{3,3}=D_{3,1}+D_{3,2}-D_{3,3}-D_{3,4}\nonumber\\
		&&\mathscr{\tilde{D}}_{3,4}=D_{3,1}-D_{3,2}-D_{3,3}-D_{3,4}
	\end{eqnarray}

	Now, invoking the reproducibility condition given by Eq.~(\ref{blom}) it follows that
	\begin{eqnarray}
		\label{trilocal31}
		\mathcal{J}_{3,1}&=&\iiint d\lambda_{1} d\lambda_{2}  d\lambda_{3} \ \mu (\lambda_{1})   \mu (\lambda_{2})  \mu (\lambda_{3})  \ \qty[ \langle A_{3,1}\rangle_{\lambda_{1}}+\langle A_{3,2}\rangle_{\lambda_{1}}] \nonumber\\
		&& \ \ \ \ \ \ \times \ \ \qty[\langle B_{3,1}\rangle_{\lambda_{1},\lambda_{2},\lambda_{3}}] \ \ \qty[ \langle C_{3,1}\rangle_{\lambda_{2}}+\langle C_{3,2}\rangle_{\lambda_{2}}+\langle C_{3,3}\rangle_{\lambda_{2}}] \nonumber\\
		&& \ \ \ \ \ \ \times \ \ \qty[\langle D_{3,1}\rangle_{\lambda_{3}}+\langle D_{3,2}\rangle_{\lambda_{3}}+\langle D_{3,3}\rangle_{\lambda_{3}}+\langle D_{3,4}\rangle_{\lambda_{3}}]   
	\end{eqnarray} 
	Since $|\langle B_{3,1}\rangle_{\lambda_{1},\lambda_{2},\lambda_{3}}|\leq1$, we obtain
	\begin{eqnarray}\label{t3cond}
		\qty|\mathcal{J}_{3,1}|&=&\iiint d\lambda_{1} d\lambda_{2}  d\lambda_{3} \ \mu (\lambda_{1})   \mu (\lambda_{2})  \mu (\lambda_{3})  \ \qty| \langle A_{3,1}\rangle_{\lambda_{1}}+\langle A_{3,2}\rangle_{\lambda_{1}}| \nonumber\\
		&& \ \ \ \ \ \ \times \ \ \qty| \langle C_{3,1}\rangle_{\lambda_{2}}+\langle C_{3,2}\rangle_{\lambda_{2}}+\langle C_{3,3}\rangle_{\lambda_{2}}| \nonumber\\
		&& \ \ \ \ \ \ \times \ \ \qty|\langle D_{3,1}\rangle_{\lambda_{3}}+\langle D_{3,2}\rangle_{\lambda_{3}}+\langle D_{3,3}+\langle D_{3,4}\rangle_{\lambda_{3}}| 
	\end{eqnarray} 
	
	The terms $|\mathcal{J}_{3,2}|$, $|\mathcal{J}_{3,3}|$, and $|\mathcal{J}_{3,4}|$ can also be written in a similar manner as Eq. (\ref{t3cond}).  Then, we obtain the following
	\begin{equation}\label{t3b}
		(\mathcal{T}_{3})_{tl}\leq \qty(\iiint d\lambda_{1} d\lambda_{2} d\lambda_{3} \ \mu(\lambda_{1}) \mu(\lambda_{2}) \mu(\lambda_{3}) \ 
		\eta_{1} \eta_{2} \eta_{3})^{\frac{1}{3}}
	\end{equation}
	where  $\eta_{1}=\big[|\langle{A_{3,1}}\rangle_{\lambda_{1}}+\langle{A_{3,2}}\rangle_{\lambda_{1}}|+|\langle{A_{3,2}}\rangle_{\lambda_{1}}+\langle{A_{3,3}}\rangle_{\lambda_{1}}|+|\langle{A_{3,3}}\rangle_{\lambda_{1}}+\langle{A_{3,4}}\rangle_{\lambda_{1}}|+|\langle{A_{3,4}}\rangle_{\lambda_{1}}-\langle{A_{3,1}}\rangle_{\lambda_{1}}|\big]$, \  $\eta_{2}=\big[|\langle{C_{3,1}}\rangle_{\lambda_{2}}+\langle{C_{3,2}}\rangle_{\lambda_{2}}+\langle{C_{3,3}}\rangle_{\lambda_{2}}|+|\langle{C_{3,1}}\rangle_{\lambda_{2}}+\langle{C_{3,2}}\rangle_{\lambda_{2}}-\langle{C_{3,3}}\rangle_{\lambda_{2}}|+|\langle{C_{3,1}}\rangle_{\lambda_{2}}-\langle{C_{3,2}}\rangle_{\lambda_{2}}+\langle{C_{3,3}}\rangle_{\lambda_{2}}|+|-\langle{C_{3,1}}\rangle_{\lambda_{2}}+\langle{C_{3,2}}\rangle_{\lambda_{2}}+\langle{C_{3,3}}\rangle_{\lambda_{2}}| \big]$, and $\eta_{3}=\big[|\langle D_{3,1}\rangle_{\lambda_{3}}+\langle D_{3,2}\rangle_{\lambda_{3}}+\langle D_{3,3}\rangle_{\lambda_{3}}+\langle D_{3,4}\rangle_{\lambda_{3}}|+|\langle D_{3,1}\rangle_{\lambda_{3}}+\langle D_{3,2}\rangle_{\lambda_{3}}+\langle D_{3,3}\rangle_{\lambda_{3}}-\langle D_{3,4}\rangle_{\lambda_{3}}|+|\langle D_{3,1}\rangle_{\lambda_{3}}+\langle D_{3,2}\rangle_{\lambda_{3}}-\langle D_{3,3}\rangle_{\lambda_{3}}-\langle D_{3,4}\rangle_{\lambda_{3}}|+|\langle D_{3,1}\rangle_{\lambda_{3}}-\langle D_{3,2}\rangle_{\lambda_{3}}-\langle D_{3,3}\rangle_{\lambda_{3}}-\langle D_{3,4}\rangle_{\lambda_{3}}|\big]$. 
	
	Since all the observables are dichotomic with eigenvalues $\pm1$, it is straightforward to derive that $\eta_{1}\leq 6$, $\eta_{2}\leq 6$, and $\eta_{3}\leq 8$. Therefore, from Eq.~(\ref{t3b}), integrating over $\lambda_{1}$, $\lambda_{2}$ and $\lambda_{3}$ we get 
	\begin{equation}\label{S3f}
		(\mathcal{T}_{3})_{tl} \leq 2 \left(6\right)^{\frac{2}{3}} \approx 6.60
	\end{equation}
	i.e., the trilocal bound $ (\mathcal{T}_{3})_{tl}=2 \left(6\right)^{\frac{2}{3}} \approx 6.60$. There are suitable states and observables for which the trilocal bound can be violated in quantum theory. Now, in the following, we evaluate the quantum optimal bound of $\mathcal{T}_{3}$.
	
	\subsection{Optimal quantum bound of the asymmetric trilocality inequality for scenario-I}\label{sost3}
	
	Here, by invoking the SOS approach, we evaluate  the optimal quantum bound of $\mathcal{T}_{3}$ without assuming the dimension of the system. We first show that there exists a positive semi-definite operator $\Gamma_{3}$ satisfying $\langle\Gamma_{3}\rangle=\zeta_3-\left(\mathcal{T}_{3}\right)_{Q}$. The existence of such operator can be proved by considering a set of operators $L_{3,j} \ \forall j\in \{1,2,3,4\}$ such that
	\begin{equation}
		\label{gamma3t1}
		\langle\Gamma_{3}\rangle=\sum\limits_{j=1}^4\dfrac{\left(\omega_{3,j}\right)^{\frac{1}{3}}}{2} \qty|L_{3,j}\ket{\psi}|^2
	\end{equation}
	where $\omega_{3,j}\geq0$ and $\omega_{3,j}=\omega^{A}_{3,j}\cdot\omega^{C}_{3,j}\cdot\omega^{D}_{3,j}$. We choose $L_{3,j}$ and $\omega_{3,j}$ as the following way

	\begin{eqnarray}
		&&\qty|L_{3,j}\ket{\psi}|=\qty| \left(\frac{\mathscr{\tilde{A}}_{3,j}}{\omega_{3,j}^{A}}\otimes\frac{\mathscr{\tilde{C}}_{3,j}}{\omega_{3,j}^{C}}\otimes\frac{\mathscr{\tilde{D}}_{3,j}}{\omega_{3,j}^{D}}\right)\ket{\psi}|^{\frac{1}{3}}-\qty|B_{3,j}\ket{\psi}|^{\frac{1}{3}}\label{L3}\\
		&& \omega_{3,j}^A=||\mathscr{\tilde{A}}_{3,j}\ket{\psi}||_2 \ ; \ \ \omega_{3,j}^C=||\mathscr{\tilde{C}}_{3,j}\ket{\psi}||_2 \ ; \ \ \omega_{3,j}^D=||\mathscr{\tilde{D}}_{3,j}\ket{\psi}||_2 \nonumber \label{wl3}\\
	\end{eqnarray}

Now, putting $\qty|L_{3,j}\ket{\psi}|$ from Eq.~(\ref{L3}) into the Eq. (\ref{gamma3t1}), after a simple algebraic evaluation, we obtain $\langle \Gamma_3 \rangle =\sum\limits_{j=1}^3 \qty(\omega_{3,j})^{\frac{1}{3}}-(\mathcal{T}_3)_Q$. Then, it follows that the quantum optimal value corresponds $\langle \Gamma_3 \rangle=0$. Therefore,
	\begin{eqnarray}\label{t3opt}
		\left(\mathcal{T}_{3}\right)_{Q}^{opt} &=&\sum\limits_{j=1}^4 \qty(\omega_{3,j}^{A}\cdot\omega_{3,j}^{C} \cdot\omega_{3,j}^{D})^{\frac{1}{3}}
	\end{eqnarray}
	
Such optimal quantum value will occur under the following optimisation condition
	\begin{equation}\label{impn3}
		L_{3,j} \ \ket{\psi}=0 \ \ \forall j \in \{1,2,3,4\}
	\end{equation}
	
	Hence, from Eq. (\ref{t3opt}) and by using the inequalities given in footnotes (\ref{f2}) and (\ref{f3}), we obtain the quantum optimal value as follows
	\begin{eqnarray}\label{tbs1}
		\left(\mathcal{T}_{3}\right)_{Q}^{opt} =2 \  \qty[\prod_{k=A,C,D}\qty(\max \sum_{j=1}^{4}\qty( \omega^{k}_{3,j})^2)] ^{\frac{1}{6}}
	\end{eqnarray}
	
 Note that the quantities $\sum_{j=1}^{3}\left( \omega^{A}_{3,j}\right)^2\leq4(2+\sqrt{2})$ and $\sum_{j=1}^{3}\left( \omega^{C}_{3,j}\right)^2\leq12$ have already derived in Eqs.~(\ref{ps2}) and (\ref{ps2b}). Now, we evaluate the quantity $\sum_{j=1}^{4}\left( \omega^{D}_{3,j}\right)^2$ from Eq.~(\ref{wl3}) as follows
	\begin{eqnarray}\label{wdt1}
		\sum\limits_{j=1}^{4}\qty(\omega^{D}_{3,j})^2&=&\bra{\psi}(16+ 2\big( \{D_{3,1},(D_{3,2}-D_{3,4})\}\nonumber\\
		&& \ \ \ \ + \{D_{3,3},(D_{3,2}+D_{3,4})\}\big)\ket{\psi}
	\end{eqnarray}
Since the quantities $D_{3,1}$ and $D_{3,3}$ appeared independently with $\{D_{3,1},(D_{3,2}-D_{3,4})\}$ and $\{D_{3,3},(D_{3,2}+D_{3,4})\}$ respectively, without loss of generality, we chose $D_{3,1}=(D_{3,2}-D_{3,4})/\nu_1$ and $D_{3,3}=(D_{3,2}+D_{3,4})/\nu_1$, where $\nu_1=|| (D_{3,2}-D_{3,4}) \ket{\psi}||_2$ and $\nu_2=|| (D_{3,2}+D_{3,4}) \ket{\psi} ||_2$. Therefore, the above Eq.~(\ref{wdt1}) reduces to
	\begin{equation}
		\label{wdt11}
		\sum\limits_{j=1}^{4}\bigg(\omega^{D}_{4,j}\bigg)^2=16+4\left[\sqrt{2-\langle\{D_{3,2},D_{3,4}\}\rangle}+\sqrt{2-\langle \{D_{3,2},D_{3,4}\}\rangle}\right] 
	\end{equation}
	The maximum value of $\sum\limits_{j=1}^{4}\qty(\omega^{D}_{4,j})^2=8(2+\sqrt{2})$ is then achieved when $\{D_{3,2},D_{3,4}\}=0$ which automatically implies $\nu_1=\nu_2=\sqrt{2}$. Therefore, for the optimal quantum violation, the linear constraints on Diana's observables are given as
	\begin{eqnarray}
		D_{3,4}-\sqrt{2} \ D_{3,1}-D_{3,2}=   D_{3,2}+D_{3,4}-\sqrt{2} \ D_{3,3}=0 \label{d44}
	\end{eqnarray}
	
Therefore, we obtain the quantum optimal value as given by
	\begin{equation}\label{Bt1}
		\left(\mathcal{T}_{3} \right)^{opt}_{Q} = 4 \ \qty[ 2\sqrt{3}\qty(1+\sqrt{2})]^{\frac{1}{3}}	\approx 8.12
	\end{equation} 
	
	It is important to remark here that the optimal quantum value $\left(\mathcal{T}_{3}\right)_{Q}^{opt}= 4 \ \qty[ 2\sqrt{3}\qty(1+\sqrt{2})]^{\frac{1}{3}}\approx 8.12$ is evaluated without specifying the dimension of both the system and observables. The states, and the observables for which the optimal value will be achieved are given in the following.
	
	\subsection{The state and observables for the optimal quantum violation of $(\mathcal{T})_{3}$}\label{st3s2}

	We further obtain relationships between the observables of all the parties in terms of the anti-commuting relations for achieving the optimal quantum violation.  The anti-commutations relations for Alice's and Charlie's observables are evaluated from Eqs. (\ref{a44}) and (\ref{cs2cond}).  The anti-commutations relations for Diana's observables are evaluated from Eq. (\ref{d44}). All the relations are given as follows 
	\begin{eqnarray}\label{obst1cond}
		&&\{A_{3,1},A_{3,2}\}=\{A_{3,2},A_{3,3}\}=\{A_{3,3},A_{3,4}\}=-\{A_{3,1},A_{3,4}\}=\sqrt{2} \  \mathbb{I}_{d} \nonumber\\
		&&\{A_{3,1},A_{3,3}\}=\{A_{3,2},A_{3,4}\}=0; \ \{D_{3,1},D_{3,3}\}=\{D_{3,2},D_{3,4}\}=0  \nonumber\\
		&&\{D_{3,1},D_{3,2}\}=\{D_{3,2},D_{3,3}\}=\{D_{3,3},D_{3,4}\}=-\{D_{3,1},D_{3,4}\}=\sqrt{2} \  \mathbb{I}_{d} \nonumber\\
		&&\{C_{3,1},C_{3,2} \}=\{C_{3,2},C_{3,3} \}=-\{C_{3,1},C_{3,3}\}=\mathbb{I}_{d} 
	\end{eqnarray}
	By using the above relations between the observables  given by Eq.~(\ref{obst1cond}) on the observables, one can always construct a set of observables for Alice and Charlie in the Hilbert space dimension $\mathcal{H}^d , \ \ \forall d\geq2$.
	
	Next, we recall the optimization condition obtained in the SOS method from the Eq.~(\ref{impn3}) to find the constraints on Bob's observable. The specific condition $L_{3,j} \ket{\psi} =0 \ \ \forall j\in\{1,2,3,4\}$ implies the following
	\begin{eqnarray} \label{bt1cond}
		B_{3,j}&=&\frac{\mathscr{\tilde{A}}_{3,j}}{\omega_{3,j}^{A}}\otimes\frac{\mathscr{\tilde{C}}_{3,j}}{\omega_{3,j}^{C}} \otimes\frac{\mathscr{\tilde{D}}_{3,j}}{\omega_{3,j}^{C}}
	\end{eqnarray}
	We explicitly construct a set of observables of Alice, Charlie, and Diana for the Hilbert space dimension $\mathcal{H}^2$ are the following.
	\begin{eqnarray}\label{tobs3}
		A_{3,1}&=& r \ \sigma_{x} + \sqrt{1-r^2}\sigma_{z} \ \ ; \ \ A_{3,2}= t \ \sigma_{x} + \sqrt{1-t^2}\sigma_{z}\nonumber\\
		\ A_{3,3}&=&t\ \sigma_{x} -\sqrt{1-t^2}\ \sigma_{z} \ \ ; \ \ A_{3,4}= r \ \sigma_{x} - \sqrt{1-r^2} \ \sigma_{z}\nonumber \ \  \\
		D_{3,1}&=& -t \ \sigma_{x} + \sqrt{1-t^2}\sigma_{z} \ \ ; \ \ D_{3,2}= -t \ \sigma_{x} - \sqrt{1-t^2}\sigma_{z}\nonumber\\
		\ D_{3,3}&=&-r\ \sigma_{x} -\sqrt{1-r^2}\ \sigma_{z} \ \ ; \ \ D_{3,4}= r \ \sigma_{x} - \sqrt{1-r^2} \ \sigma_{z}\nonumber \ \  \\
		C_{3,1}&=&\sigma_{x} \ \ ; \ \  C_{3,2}=\sigma_{y} \ \ ; \ \ C_{3,3}=\sigma_{z} 
	\end{eqnarray}
	where $r=\frac{1}{2}\sqrt{2-\sqrt{2}}$ and $t=\frac{1}{2}\sqrt{2+\sqrt{2}}$.
	Note that employing the above-mentioned observables, we find that the quantum optimal value $\left(\mathcal{T}_{3}\right)_{Q}^{opt}\approx7.23$ is achieved when three maximally entangled two-qubit states are shared between Alice-Bob, Charlie-Bob and Diana-Bob.


	\section{Asymmetric trilocal network: scenario-II}\label{secVI}
	
	\begin{figure}[ht]
		\centering
		\includegraphics[width=0.6\linewidth]{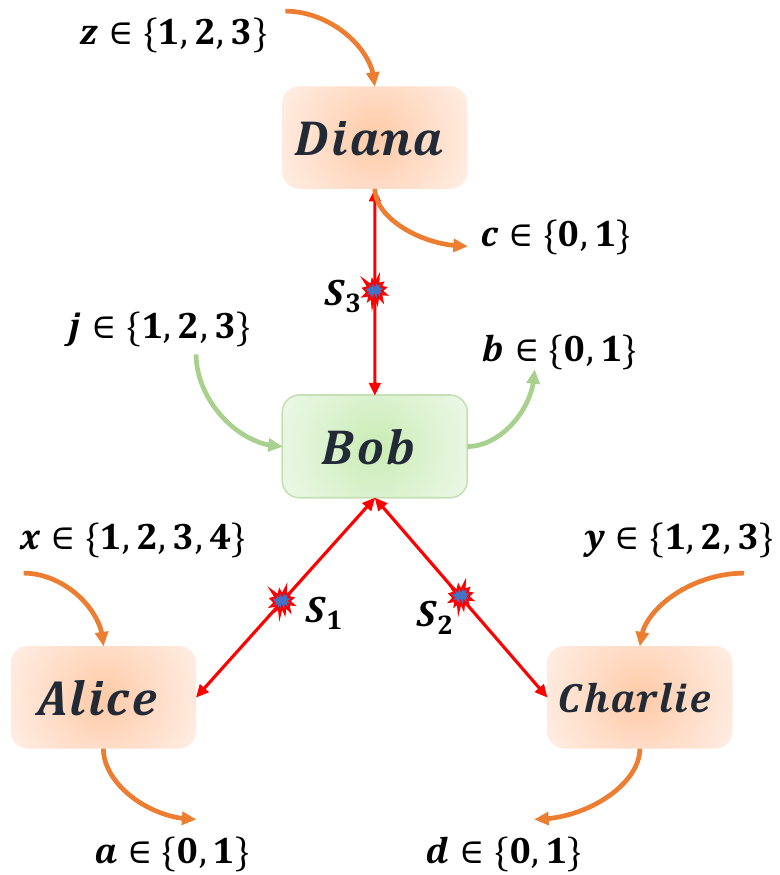}
		\caption{Asymmetric trilocal network scenario featuring three edge parties (Alice, Charlie, and Diana) and a central party Bob. The independent sources $S_{1}$, $S_{2}$ and $S_{3}$ emit physical systems for Alice-Bob, Charlie-Bob, and Diana-Bob respectively.}
		\label{t1s2}
	\end{figure}
	
	Here we present another asymmetric trilocal network scenario in which one of the edge party Alice performs one of four dichotomic measurements and rest of the edge parties Charlie and Diana as well as the central party Bob perform three dichotomic measurements. In this scenario, let us consider the nonlinear trilocal inequality of the following form
	\begin{equation}
		\label{t3s2}
		\mathcal{T}_{3}^{\prime}=\sum_{j=1}^{3}\qty|\ \Big\langle \mathscr{\tilde{A}}_{3,j}^{\prime} B_{3,j}\mathscr{\tilde{C}}_{3,j}^{\prime} \mathscr{\tilde{D}}_{3,j}^{\prime}\Big\rangle \ |^{\frac{1}{3}} \ \leq  \ (\mathcal{T}_{3}^{\prime})_{tl}
	\end{equation}	
where $(\mathcal{T}_{3}^{\prime})_{tl}$ is the local bound of $\mathcal{T}_{3}^{\prime}$ and the quantities $\mathscr{\tilde{A}}_{3,1}^{\prime}=A_{3,1}+A_{3,2}+ A_{3,3}-A_{3,4}$; $\mathscr{\tilde{D}}_{3,1}^{\prime}=D_{3,1}+D_{3,2}+D_{3,3}$; $\mathscr{\tilde{A}}_{3,2}^{\prime}=A_{3,1}+A_{3,2}- A_{3,3}+A_{3,4}$; $\mathscr{\tilde{D}}_{3,2}^{\prime}=D_{3,1}-D_{3,2}+D_{3,3}$; $\mathscr{\tilde{A}}_{3,3}^{\prime}=A_{3,1}-A_{3,2}+ A_{3,3}+A_{3,4}$; $\mathscr{\tilde{D}}_{3,3}^{\prime}=D_{3,1}-D_{3,2}-D_{3,3}$; $\mathscr{\tilde{C}}_{3,1}^{\prime}=C_{3,j}+C_{3,j+1}$ satisfying $C_{3,4}=-C_{3,1}$. We find (see appendix \ref{apt21}) that the trilocal bound in this scenario is given by
	\begin{equation}
		(\mathcal{T}_{3}^{\prime})_{tl}= 2(15)^{\frac{1}{3}} \approx 4.93
	\end{equation}	
	It can be shown that there are suitable states and observables for which quantum correlations violate the trilocal bound. In the following, we evaluate the quantum optimal bound of $\mathcal{T}_{3}^{\prime}$.

	
	The optimal quantum bound of $\mathcal{T}_{3}^{\prime}$ without assuming the dimension of the system is derived by suitably invoking the SOS approach as discussed earlier. The explicit construction of the SOS method is provided in Appendix (\ref{apt22}). The optimal quantum bound is found to be 
	\begin{eqnarray}
		\qty(\mathcal{\mathcal{T}}_{3}^{\prime})_{Q}^{opt}=6
	\end{eqnarray} 
	
	It is to be noted here that the obtained optimal quantum bound $\qty((\mathcal{\mathcal{T}}_{3}^{\prime})_{Q}^{opt}= 6)$ in this scenario is lesser than that obtained $\qty((\mathcal{\mathcal{T}}_{3})_{Q}^{opt} \approx7.23)$ in the earlier scenario-I. Thus, the trilocal as well as the quantum optimal bound depends on how the asymmetry invoked in the trilocal scenario. 
	
	The corresponding state and observables for which the quantum optimal bound is achieved is given in Appendix (\ref{apt23}).


\section{Asymmetric Network for arbitrary $n$} \label{secforn}

	Let us now provide a sketch of the result in an asymmetric bilocal network scenario for arbitrary $n$. 
	Alice performs one of $m_A=2^{n-1}$ dichotomic measurements and both the central party (Bob) and Charlie perform one of $m_B=n$ and $m_c=n$ dichotomic measurements respectively. In this scenario, we consider the non-linear asymmetric bilocality inequality for arbitrary input $n$ of the following form
	\begin{equation}
		\label{sn}
		\mathcal{B}_{n}=\sum_{j=1}^{n}\sqrt{|\mathcal{I}_{n,j}|} \ \leq \ \left(\mathcal{B}_{n}\right)_{bl}
	\end{equation}
	where, $\left(\mathcal{B}_{n}\right)_{bl}$ is the bilocal bound of $\mathcal{B}_{n}$  and $\mathcal{I}_{n,j}=\Big\langle\tilde{\mathcal{A}}_{n,j}\otimes B_{n,j}\otimes\tilde{\mathcal{C}}_{n,j}\Big\rangle$ with $\tilde{\mathcal{A}}_{n,j}$ and $\tilde{\mathcal{C}}_{n,j}$ are unnormalised observables given as follows
	\begin{eqnarray}\label{tilde2}
		\tilde{\mathcal{A}}_{n,j}=\sum\limits_{x=1}^{2^{n-1}}(-1)^{y^{x}_{j}} A_{n,x} \ ; \ \ \tilde{\mathcal{C}}_{n,j}=\sum_{z=1}^{n}\left(C_{n,z}+C_{n,z+1}\right)
	\end{eqnarray}
	with $C_{n,n+1}=-C_{n,1}$ and $y^{x}_{j}\in \{0,1\} \ \forall j\in[n]$. For our purpose, by using  the encoding scheme which was earlier introduced in the context of Random Access Codes (RACs) \cite{Ambainis2008, Ghorai2018, Pan2020, Munshi2021, Munshi2022} protocol, we fix the values of $y^{x}_{j}$ in the following way. Let us consider a random variable $y^{\alpha}\in \{0,1\}^{n}$ with $\alpha\in \{1,2...2^{n}\}$. Each element of the bit string can be written as $y^{\alpha}=y^{\alpha}_{j=1} y^{\alpha}_{j=2} y^{\alpha}_{j=3} .... y^{\alpha}_{j=n}$. For  example, if $y^{\alpha} = 011...00$ then $y^{{\alpha}}_{j=1} =0$, $y^{{\alpha}}_{j=2} =1$, $y^{{\alpha}}_{j=3} =1$ and so on. We  denote  the $n$-bit binary strings as $y^{x}$. Here we consider the bit strings such that for any two $x$ and $x'$,  $y^{x}\oplus_2y^{x'}=11\cdots1$. Clearly, we have $x\in \{1,2...2^{n-1}\}$ constituting the inputs for Alice. If $x=1$, we get all the first bit of each  bit  string $y_j$ for every $j\in \{1,2 \cdots n\}$.
	
	Now it follows from the reproducibility condition given by Eq.~(\ref{blom}) that
	\begin{equation}
		\label{bilocaln1}
		\mathcal{I}_{n,1}=\iint \mu (\lambda_{1}) \mu (\lambda_{2}) d\lambda_{1} d\lambda_{2} \langle \tilde{\mathcal{A}}_{n,1}\rangle_{\lambda_{1}} \langle B_{n,1}\rangle_{\lambda_{1},\lambda_{2}} \langle \tilde{\mathcal{C}}_{n,1}\rangle_{\lambda_{1}}
	\end{equation} 
	
	Since $|\langle B_{n,1}\rangle_{\lambda_{1},\lambda_{2}}|\leq1$, we obtain from the above Eq. (\ref{bilocaln1})
	\begin{eqnarray}\label{ncond}
		|\mathcal{I}_{n,j}|&\leq& \iint \mu (\lambda_{1}) \mu (\lambda_{2}) d\lambda_{1} d\lambda_{2}   \  \ \Big|  \big \langle \tilde{\mathcal{A}}_{n,j}\big \rangle_{\lambda_{1}}\Big| \ \  \Big|  \langle \tilde{\mathcal{C}}_{n,j}\rangle_{\lambda_{2}}\Big|\nonumber\\
		&&\ \ \ \ ; \forall j \in \{1,2,...,n\}
	\end{eqnarray}

	By putting the values of Eq. (\ref{ncond}) in Eq. (\ref{sn}) and, then by applying the property of the inequality given in footnote \ref{f2}, we obtain
	\begin{equation}\label{Sn}
		\left(\mathcal{B}_{n}\right)_{bl}\leq\sqrt{\int d\lambda_{1}\mu(\lambda_{1}) \ \big|  \big \langle \tilde{\mathcal{A}}_{n,j}\big \rangle_{\lambda_{1}}\big|} \times\sqrt{\int d\lambda_{2}\mu(\lambda_{2}) \ \big|  \big \langle \tilde{\mathcal{C}}_{n,j}\big \rangle_{\lambda_{2}}\big|}
	\end{equation}
	
	Note that all the observables $A_{n,x}$ and $C_{n,z}$ are dichotomic with eigenvalues $\pm1$ . In \cite{Munshi2021, Munshi2022} it was derived that the  value of $\big|\big \langle \tilde{\mathcal{A}}_{n}\big \rangle_{\lambda_{1}}\big|=n \ \binom{n-1}{\lfloor \frac{n-1}{2}\rfloor}$ and $\big|\big \langle \tilde{\mathcal{C}}_{n}\big \rangle_{\lambda_{2}}\big|=(2n-2)$. Hence, the bilocal bound is given by
	\begin{equation}
		\left(\mathcal{B}_{n}\right)_{bl}= \sqrt{ 2n(n-1) \  \binom{n-1}{\lfloor\frac{n-1}{2}\rfloor}}
	\end{equation} 		
where $\lfloor x \rfloor$ denotes the largest integer smaller or equal to $x$.

	We derive the optimal quantum value of $\left(\mathcal{B}_{n}\right)_{Q}$ of the bilocality inequality proposed in Eq. (\ref{sn}) by again invoking the SOS approach. We consider a positive semi-definite operator $\langle\gamma_{n}\rangle\geq 0$, that can be expressed as $\langle \gamma_{n}\rangle=\beta_{n}\mathbb{I}-\left(\mathcal{B}_{n}\right)_{Q}$. Where $\beta_{n}$ is the optimal value that can be obtained when $\langle \gamma_{n}\rangle$ is equal to zero. This can be proved by considering a set of positive operators $M_{n,j}$ which is polynomial functions of   $A_{n,x}$, $C_{n,z}$,  $B_{n,j}$ such that 	
	
	\begin{equation}\label{gamman}
		\langle\gamma_{n}\rangle=\sum_{j=1}^{n}\dfrac{(\omega_{n,j})^{\frac{1}{2}}}{2}\langle\psi|(M_{n,j})^{\dagger}(M_{n,j})|\psi\rangle
	\end{equation}	
	where $\omega_{n,j}$ is suitable positive numbers and $\omega_{n,j}=\qty(\omega^{A}_{n,j}) \ \qty(\omega^{C}_{n,j})$. The optimal quantum value of $\left(\mathcal{B}_{n}\right)_{Q}$ is obtained if $\langle \gamma_{n}\rangle=0$, implying that $M_{n,j}|\psi\rangle=0$. 
	We choose a  suitable set of  positive operators $M_{n,j} $ (with $j\in [n]$), such that
	
	\begin{eqnarray}\label{mnj}
		M_{n,j}|\psi\rangle&=&\sqrt{\bigg|\left(\mathcal{A}_{n,j}\otimes\mathcal{C}_{n,j}\right) |\psi\rangle
			\bigg|} -\sqrt{\left|B_{n,j}|\psi\rangle\right|}
	\end{eqnarray}
	where $\mathcal{A}_{n,j}=\frac{\tilde{\mathcal{A}}_{n,j}}{\omega^{A}_{n,j}}$ and $\mathcal{C}_{n,j}=\frac{\tilde{\mathcal{C}}_{n,j}}{\omega^{C}_{n,j}}$ with $\omega_{n,j}^{A}=||\tilde{\mathcal{A}}_{n,j}|\psi\rangle||_2$ and $\omega_{n,j}^{C}=||\tilde{\mathcal{C}}_{n,j}|\psi\rangle||_2$. By inserting Eq. (\ref{mnj}) in Eq. (\ref{gamman}), we obtain 
	$\langle\gamma_{n}\rangle=-\left(\mathcal{B}_{n}\right)_{Q} +\sum\limits_{j=1}^{n}\left(\omega_{n,j}\right)^{\frac{1}{2}}$. 	The optimal value of $\left(\mathcal{B}_{n}\right)_{Q}$ is obtained if  $\langle \gamma_{n}\rangle=0$. Therefore the optimal value is given as follows
	
	\begin{eqnarray}\label{snm}
		\left(\mathcal{B}_{n}\right)_{Q}^{opt} =\max\left(\sum\limits_{j=1}^{n}(\omega_{n,j})^{\frac{1}{2}}\right)
	\end{eqnarray}
	
	Now, by using the inequality given in footnote \ref{f2} along with the convex inequality given in the footnote \ref{f3}, we obtain
	\begin{eqnarray}\label{snm1}
		\left(\mathcal{B}_{n}\right)_{Q}^{opt} = \max\left[\prod_{k=A,C}^{}\left(n\sum_{j=1}^{n}\left( \omega^{k}_{n,j}\right)^2\right)^{\frac{1}{2}} \right] ^{\frac{1}{2}}
	\end{eqnarray}
	
We can always evaluate each term $(\omega_{n,j}^{A})^2$ and $(\omega_{n,j}^{C})^2$ by following the procedure discussed for $n=3$ case. However, such evaluation will take rigorous algebraic treatment which we are skipping here. Thus, instead of following the same path, in the following, we present an argument that leads to the optimal quantum value of $(\mathcal{B}_n)_Q$.

To begin with, let us revisit the expression $\mathcal{B}_{n}$ given by Eq.~(\ref{sn}). Since all $\mathcal{I}_{n,j}$ are real numbers and positive by construction, we can always invoke the convex inequality (as mentioned in the footnote \ref{f3}) to obtain the following
\begin{equation}
(\mathcal{B}_{n})_Q \leq \left(n\sum\limits_{j=1}^n \mathcal{I}_{n,j}\right)^{\frac{1}{2}}
\end{equation}
Then the quantity $\sum\limits_{j=1}^n \mathcal{I}_{n,j}$ is evaluated from the Eq.~(\ref{Inj}).
\begin{equation}\label{arg1}
\sum\limits_{j=1}^n \mathcal{I}_{n,j}= \left\langle\sum\limits_{j=1}^n \left(\tilde{\mathcal{A}}_{n,j} \otimes B_{n,j} \otimes \tilde{\mathcal{C}}_{n,j}\right) \right\rangle
\end{equation}  
	
	Now, by noting the optimization condition $M_{n,j}|\psi\rangle=0$ form the SOS method, we write
	\begin{equation}\label{bobn1}
		B_{n,j}=\mathcal{A}_{n,j} \otimes \mathcal{C}_{n,j}\ \ \ \forall j\in\{1,2,...,n\}
	\end{equation}
	
	Hence, from Eq.~(\ref{bobn1}), we conclude that for achieving the quantum optimal value, it is sufficient to assume that Bob measures his system on a product basis. Thus, Bob's observable are given by $B_{n,j}=B_{n,j}^A\otimes B_{n,j}^C$, where $B_{n,j}^A$ and $B_{n,j}^C$ are normalised observable. The ability to express Bob's observable in such a way buoyed up the fact that for the optimal value, the quantity $\sum\limits_{j=1}^n \mathcal{I}_{n,j}$ can be expressed as follows
	\begin{equation}
	\max	\sum\limits_{j=1}^n \mathcal{I}_{n,j} = \qty(\mathcal{S}_n)^{opt}_Q \  \qty(\mathcal{L}_n)^{opt}_Q	\end{equation}
	where $\mathcal{S}_n$ is the Bell functional proposed in \cite{Ghorai2018} and $\mathcal{L}_n$ is the $n$-settings Chain-Bell functional \cite{Braunstein1989}. Note that the optimal quantum values of $\mathcal{S}_n$ and $\mathcal{L}_n$ have already been derived as $\left(\mathcal{S}\right)_Q^{opt}=2^{n-1}\sqrt{n}$ and $\left(\mathcal{L}_n\right)_Q^{opt}=2n\cos\frac{\pi}{2n}$ respectively \cite{Ghorai2018, Supic2016}. By combining these results, it is straightforward to obtain the quantum optimal bound of $(\mathcal{B}_{n})_Q$.
	\begin{eqnarray}
		\qty(\mathcal{B}_{n})_{Q}^{opt} \  = \ \qty( 2^{n} \ n^{\frac{3}{2}} \
		\cos\frac{\pi}{2n})^{\frac{1}{2}}
	\end{eqnarray}
	which violates the local bound $\left(\mathcal{B}_{n}\right)$ for any arbitrary $n$.

	\subsection{The state and observables for the optimal quantum violation of $(\mathcal{B})_{n}$}\label{subn}
	
	Note that the optimal quantum value $\left(\mathcal{B}_{n}\right)_{Q}^{opt}$ is evaluated without specifying the dimension of both the system and observables. Importantly, from the argument presented in the preceding section, the optimal value is achieved when both the quantity $\mathcal{S}_n$ and $\mathcal{L}_n$ are optimized simultaneously. It has earlier been shown \cite{Ghorai2018} that optimal of $\mathcal{B}_n$ implies the observables $B_{n,j}^A$ are mutually anti-commuting, i.e., $\{B_{n,j}^A,B_{n,i}^A\}=0 \ \forall i,j$. 
	
Thus, it is crucial to remark that to achieve the optimal quantum violation of the asymmetric bilocality inequality, there should be $n$ mutually anti-commuting operators. 
Further it follows that $\mathcal{A}_{n,j} \in \mathcal{H}^d_{A}$ with necessarily $(d_{A})_{min}=2^{\lfloor n/2\rfloor}$. Thus, the Bob's observables $B_{n,j}=\mathcal{A}_{n,j} \otimes \mathcal{C}_{n,j}$ must belong to $\mathcal{H}^d_{B}$ with necessarily $(d_{B})_{min}>2^{\lfloor n/2\rfloor}$. We can also conclude that the optimal value cannot be achieved if Alice-Bob and Bob-Charlie share a single copy of the maximally entangled two-qubit state. 
	
	Hence, taking a cue from the preceding discussions, we find that if there should be \textit{at least} $N=\lfloor n/2\rfloor$ copies of maximally entangled two-qubit states between Alice-Bob and \textit{at least} a single copy of maximally entangled two-qubit state between Bob-Charlie, then the quantum optimal value $\left(\mathcal{B}_{n}\right)_{Q}^{opt}$ will achieve for the observables in total dimension  $d_{min} = 2^N \times 2^{N+1}\times 2=4^{N+1}$.

Now, following a similar argument, it is straightforward to obtain the bilocal as well as the quantum optimal bound of asymmetric bilocality inequality for Scenario-2 where Alice and Bob perform $2^{n-1}$ number of measurements and Charlie performs $n$ number of measurements. In this scenario, the bilocal bound of the $n$-settings asymmetric bilocal inequality is given by 
\begin{equation}
    \mathcal{B}_n^{\prime}=\sum\limits_{j=1}^{2^{n-1}} \sqrt{\qty|\Big\langle\tilde{\mathcal{A}}_{n,j}^{\prime}\otimes B_{n,j}\otimes\tilde{\mathcal{C}}_{n,j}^{\prime}\Big\rangle|} \leq \sqrt{ 2n(2^{n-1}-1) \ \binom{n-1}{\lfloor\frac{n-1}{2}\rfloor}}
\end{equation}
where $\tilde{\mathcal{A}}_{n,j}=\sum\limits_{x=1}^{2^{n-1}}\left(C_{n,z}+C_{n,z+1}\right)$ with $A_{n,n+1}=-A_{n,1}$ and $\tilde{\mathcal{C}}_{n,j}=\sum_{z=1}^{n}(-1)^{y^{z}_{j}} C_{n,z}$. The optimal quantum bound in this case is given by 
\begin{equation}
    (\mathcal{B}_n)_Q^{opt} \ = \ \qty(2^{2n-1} \ \sqrt{n}\ \cos\frac{\pi}{2^{n}})^{\frac{1}{2}}
\end{equation}

Next, for the asymmetric trilocal network scenario-I, Alice, Bob and Diana perform $2^{n-1}$ measurements and Charlie performs $n$ measurements. In this scenario, the trilocal bound is given by		
\begin{equation}
(\mathcal{T}_n)_{tl}=  \qty[2n \ (2^{n-1}-1)\ \binom{n-1}{\lfloor\frac{n-1}{2}\rfloor} \ \left\lfloor 2^{2n-3}+\frac{1}{2}\right\rfloor]^{\frac{1}{3}}
\end{equation} 

The optimal quantum value can be written in the form $(\mathcal{T}_n)_{tl}^{opt}=(\mathcal{L}_{2^{n-1}})^{opt}_Q \ (\mathcal{G}_{2^{n-1}})^{opt}_Q \ (\mathcal{S}_{n})^{opt}_Q$. The Bell functional $\mathcal{G}_n$ is the family of n-settings Bell inequalities proposed in \cite{Gisin1999} with the optimal quantum value is given by $(\mathcal{G}_n)_Q^{opt}=2n\cos\frac{\pi}{2n}/\sin\frac{\pi}{n}$. Thus, the corresponding optimal quantum bound of $(\mathcal{T}_n)_{tl}$ is given by
\begin{eqnarray}
(\mathcal{T}_n)_Q^{opt} \ &=& 2^{n-1} \ \qty(2 \ \sqrt{n} \ \cot\frac{\pi}{2^n})^{\frac{1}{3}}
\end{eqnarray}

Now, for the asymmetric trilocal network scenario-II, the respective trilocal and optimal quantum bound is given as follows
\begin{equation}
    (\mathcal{T}_n^{\prime})_{tl}=\qty(2n(n-1)\binom{n-1}{\lfloor\frac{n-1}{2}\rfloor} \left\lfloor \frac{n^{2}+1}{2}\right\rfloor)^{\frac{1}{3}}
\end{equation}
\begin{equation}
    (\mathcal{T}_n^{\prime})_Q^{opt}= \qty(2^{n}n^{\frac{5}{2}}\cot \frac{\pi}{2n})^{\frac{1}{3}}
\end{equation}
The optimal quantum values $ (\mathcal{T}_n)_Q^{opt}>(\mathcal{T}_n)_{tl}$  and $ (\mathcal{T}_n^{\prime})_Q^{opt}>(\mathcal{T}_n^{\prime})_{tl}$ for any value of  $n$, thereby demonstrating the nonlocality in trilocal network featuring arbitrary inputs.


\section{Resistance to white noise}\label{secVII}
		
	\subsection{Resistance to white noise for bilocality scenario-I and II}\label{rbl}

	Let us assume that each of both the independent sources $S_{1}$ and $S_{2}$ does not produce maximally entangled two-qubit state, but a mixture of maximally entangled state with a white noise, known as Werner state \cite{Werner1989}. Let the two sources produce such Werner states with different noise parameters $v_1$ and $v_2$. The Werner states between Alice and Bob is $\rho^{w}_{AB}(v_1)$ and for Bob and Charlie is $\rho^{w}_{BC}(v_2)$, given by $	\rho^{w}_{AB}(v_k)=v_{k}\ket{\psi}\bra{\psi}+(1-v_{k})\frac{\mathbb{I}}{4}$ with $k\in\{1,2\}$ and $\ket{\psi}\bra{\psi}$ is a maximally entangled two-qubit state and the joint tripartite physical system is given  by  $\rho^{w}_{ABC}(v_1,v_2)=\rho^{w}_{AB}(v_1)\otimes\rho^{w}_{BC}(v_2)$. 
	
For convenience, we first evaluate the robustness of the asymmetric bilocality scenario-I for arbitrary $n$. Since the optimal quantum violation of $\mathcal{B}_{n}$ is achieved when $\lfloor n/2\rfloor$ copies of maximally entangled two-qubit states shared between Alice-Bob and a single copy of maximally entangled two-qubit state is shared between Bob-Charlie. Thus,
we take the Werner states of the form $\mathcal{W} = \rho_{AB}(v_{1})^{\otimes N} \otimes \rho_{BC}(v_{2})$ where we take $N=\lfloor n/2\rfloor$. Then, by invoking the conditions on the
observables given in Sec.~(\ref{subn}), we obtain
\begin{equation}
\qty(\mathcal{B}_n)^{\mathcal{W}}_{Q} =\qty(v_{1}^{N}v_{2})^{\frac{1}{2}} \qty( 2^{n} \ n^{\frac{3}{2}} \ \cos\frac{\pi}{2n})^{\frac{1}{2}}
\end{equation}
Therefore, in this noisy case if we take all the noise parameter is the same ($v$), the quantum violation will be achieved when 
	\begin{equation}\label{vn}
		v^{N+1} > 2^{1-n} \ \sqrt{n} \ \qty(1-\frac{1}{n}) \  \binom{n-1}{\lfloor\frac{n-1}{2}\rfloor} \ \sec\frac{\pi}{2n}
	\end{equation}
Note that $n=2,3$ correspond $N=1$. For $n=2$ the critical noise parameter for each of the Werner state is $v_c = 1/\sqrt{2}$. This is exactly as to be expected because this is the critical noise parameter for the Warner state for violation of the CHSH inequality. On the other hand, for $n=3$ the critical noise parameter $v_{c}=\sqrt{2/3} \approx 0.82$ which is greater that $1/\sqrt{2}$. So, there is no advantage has been found over the standard Bell nonlocality or bilocality scenario for demonstrating the quantum nature of the noisy maximally entangled bipartite states.
	
In the asymmetric bilocality scenario-II, for demonstrating nonlocality, the critical noise parameter is found to be
\begin{equation}
v^{N+1} > \sqrt{n} \ 2^{1-n} \ \qty(1-2^{1-n}) \  \binom{n-1}{\lfloor\frac{n-1}{2}\rfloor} \ \sec(\frac{\pi}{2^{n}})
\end{equation}

For $n=3$, the critical parameter in scenario-II is given by $v_c=\frac{3}{4}\sqrt{3-\frac{3}{\sqrt{2}}}\approx 0.84$. Thus, in the asymmetric bilocality scenario, scenario-I is more robust against the white noise than the scenario-II.

\subsection{Resistance to white noise for trilocality scenario-I and II}\label{rtl}
	
In the trilocality scenario, all the independent sources produces Werner state with noise parameters $v_1$, $v_2$, and $v_3$. The joint four-partite physical system is given  by  $\rho^{w}_{ABCD}(v_1,v_2,v_3)=\rho^{w}_{AB}(v_1)\otimes\rho^{w}_{BC}(v_2)\otimes\rho^{w}_{DB}(v_3)$. 
	
In the asymmetric trilocality scenario-I, the critical noise parameter for demonstrating nonlocality is given by	
\begin{equation}
v^{N+2}> \sqrt{n} \ 2^{2(1-n)} \ \qty(1-2^{1-n}) \ \binom{n-1}{\lfloor\frac{n-1}{2} \ \rfloor} \left\lfloor 2^{2n-3}+\frac{1}{2}\right\rfloor \ \tan (\frac{\pi}{2^n})   
\end{equation}
For $n=3$ critical noise parameter per Werner state is $v_c= \frac{\sqrt{3}}{\sqrt[3]{2 \left(\sqrt{2}+2\right)}}\approx 0.92$. 
	
On the other hand, in the asymmetric trilocality scenario-II, the critical noise parameter is given by
	\begin{equation}
	v^{N+2}> 2^{1-n} \ \frac{1}{\sqrt{n}} \ \qty(1-\frac{1}{n}) \ \binom{n-1}{\lfloor\frac{n-1}{2} \ \rfloor} \left \lfloor \frac{n^{2}+1}{2}\right\rfloor \ \tan(\frac{\pi}{2n})
	\end{equation}
For $n=3$, the critical noise parameter per Werner state is $v_c= (\frac{5}{9})^{\frac{1}{3}}\approx 0.82$. Thus, in the asymmetric trilocality cases, the scenario-II is more robust to white noise than the Scenario-I.
	
	
	\section{Summary and discussion}\label{secVIII}

In this work, we have explored the quantum nonlocality in arbitrary `$n$' input asymmetric bilocal as well as trilocal network scenario. The asymmetric bilocal scenario proposed here features two edge parties Alice and Charlie who perform $2^{n-1}$ and $n$ number of measurements, respectively.  We derive two families of bilocality inequalities specifically designed for the asymmetric scenario when the central party Bob measures $n$ (Fig. \ref{b2s1}) and $2^{n-1}$ number of measurements.
	
Furthermore, we have extended the asymmetric network scenario into the trilocal network. In particular, we have introduced two variants of asymmetric trilocal network - (i) when one edge party, Alice, performs $2^{n-1}$ measurements, the other two edge parties, Charlie and Diana perform $n$ measurements each and the central party Bob performs $2^{n-1}$ measurements. (ii) When one edge party, Charlie, performs $n$ measurements, the other two edge parties, Alice and Diana perform $2^{n-1}$ measurements each and the central party Bob performs $n$ measurements.
	
 In Secs. \ref{secIII} and \ref{secIV}, the detailed analytical treatments of the bilocal bounds for the proposed asymmetric bilocal scenarios have been provided. It has been found that the bilocal bounds depend on how the asymmetricity is invoked within the bilocal scenario. In particular, the bilocal bounds have been found to be $4.89$ and $6$ for scenarios I and II respectively.  Subsequently, by invoking the SOS technique, the DI optimal quantum bound for the asymmetric bilocality scenario-I $\Big((\mathcal{B}_{3})_{Q}^{opt}=6\Big)$ is found to be less than that obtained $\Big((\mathcal{B}_{3}^{\prime})_{Q}^{opt}=4\left[3(2+\sqrt{2})\right]^{\frac{1}{4}}\approx7.16\Big)$ in the scenario-II (Sec. \ref{sos3s1}). In both the optimization processes corresponding to scenarios I and II, we obtain the relational constraints on the observables along with shared states required for Alice-Bob and Bob-Charlie (Secs \ref{sb3s1} and \ref{aps23}). The constraints on the observables of all the parties in terms of the anti-commuting relations have also been provided in Secs.~(\ref{secIII}) and (\ref{secV}). 
	
For the asymmetric trilocality scenarios, the trilocal bounds evaluated to be approximately $6.60$ and $4.93$  for the scenarios I and II, respectively (Secs. \ref{secV} and \ref{secVI}). Consequently, using the SOS method, while for the scenario I, the optimal quantum bound is found to be $(\mathcal{T}_3)_Q^{opt}=4(2\sqrt{3}+\sqrt{6})^{\frac{1}{3}}\approx7.23$ for scenario-II, it is found to be $(\mathcal{T}_3^{\prime})_Q^{opt}=6$. In this process, we have also obtained the constraints relations of all the parties observables in terms of the anti-commutation relations (Secs. \ref{st3s2} and \ref{apt23}). We have demonstrated that the optimal quantum bound will be achieved if all the edge parties share three maximally entangled two-qubit states with the central party Bob.   
	
Moreover, we have generalised our results for arbitrary $n$. For this purpose, using the SOS method, we first established that it is sufficient for Bob to
measure in the product basis in order to obtain the optimal quantum violations in both the bilocality and trilocality scenarios. Then an interesting algebraic manipulation reduces the nonlinear bilocal inequalities into a product of two standard Bell inequalities (Sec.~\ref{secforn}). Importantly, such reduction in terms of the product of two standard Bell inequalities will only be possible at the optimal condition. It is crucial to note here that the rigorous algebraic manipulation for the case of $n = 3$ buoyed up such deep-seated  understanding of the quantum optimal bound, which then provides the necessary intuition for such a simple proof for the case of arbitrary $n$. Then, following the similar argument of the bilocality scenarios, the optimal quantum bound of asymmetric trilocality inequalities have also been evaluated.
	
Finally, for both the asymmetric bilocal and trilocal scenario, we have demonstrated the robustness of the quantum violations of the proposed inequalities in the presence of white noise. We found that the proposed asymmetric inequalities are most robust to white noise in the simplest bilocal scenario with two measurement settings for each parties. In this case, two Werner state exhibit nonlocality if $v_1 v_2 > \frac{1}{2}$. Although unfortunately, our proposed inequality becomes less robust with increasing number of measurement settings or with increasing party (or, source), it indeed possesses some independent interests.  

We conclude by raising some open questions which can be studied in future. 

i) Since the optimal quantum violations of the asymmetric bilocality inequalities cannot always be achieved with the single copy of maximally entangled two-qubit state. In fact, for the asymmetric bilocality scenario-I, while Alice-Bob needs to share at least of $\lfloor n/2 \rfloor$ copies of maximally entangled two-qubit state Bob-Charlie need to share at least one copy of it. Similar results can be proved for other scenarios also. Thus, in the network scenario, our proposed inequality has the potential to be used as a dimension witness. Construction of such proof may lead to a wide variety of interesting results in the field of self-testing of many copies of maximally entangled two-qubit states, cryptographic applications, or randomness generation protocols.
	
ii) A straightforward extension of our proposed bilocality inequality would be to invoke the 4-outcome measurement scenario for Bob. In this regard, for the symmetric bilocal scenario, different bilocal inequalities have been tailored to the $4$-outcome scenario in the context of both the Bell state measurement and the elegant joint measurement scheme  \cite{Tavakoli2021} for Bob. Although the inequality with  Bell state measurement does not provide any advantage over the usual bilocal scenario, the inequality involving elegant joint measurement is more advantageous in presence of the noise than the earlier bilocal or standard Bell inequalities. Thus, an extensive study of such a scenario in the asymmetric case may lead to interesting findings.  
	
iii) Of course, one can always generalize our asymmetric bilocal scenario by going beyond the four-party three-independent sources into a multiparty multi-source scenario. From our evaluation of the robustness to white noise, the asymmetric bilocal scenario does not provide any advantage over the noise tolerance over the standard Bell scenario. However, with increasing the number of parties and sources, one can introduce the asymmetry in many ways, which may lead to a multiparty $n$-locality inequality that may provide such advantages.  
	
In sum, the essence of our present work lies in constructing a family of asymmetric bilocality as well as trilocality inequalities and evaluating their quantum optimal bounds, importantly, by not specifying the dimension of the system or the dichotomic observables. Our work has the potential to open up interesting avenues for future research such as self-testing of many copies of entangled states, sequential sharing of quantum correlations, unbounded generation of randomness, and secret key sharing in one-to-many scenarios that calls for further study.

\section*{Acknowledgments}
S.S.M. acknowledges the UGC fellowship [Fellowship No.
16-9(June 2018)/2019(NET/CSIR)]. SS and AKP acknowledge the support from the project DST/ICPS/QuST/Theme 1/2019/4.

	\begin{widetext}
		\appendix
		
		\section{Detailed calculation for the asymmetric bilocality network scenario-II}\label{aps2}
		
		\subsection{Bilocal bound for the asymmetric bilocality network scenario-II}\label{aps21}
		\begin{equation}
			\label{n31s2a1}
			\mathcal{B}_{3}^{\prime}=\sum_{j=1}^{4}\sqrt{ \ \qty| \ \Big\langle \mathcal{\tilde{A}}_{3,j}^{\prime}B_{3,j}\mathcal{\tilde{C}}_{3,j}^{\prime}\Big\rangle \ | \ } \ \leq \  (\mathcal{B}_{3}^{\prime})_{bl}
		\end{equation}
where the quantities $\mathcal{\tilde{A}}_{3,j}^{\prime}$ and $\mathcal{\tilde{C}}_{3,j}^{\prime}$ are defined in Sec.~(\ref{secIV}). Now, invoking the reproducibility condition given by Eq.~(\ref{blom}) and since $|\langle B_{3,1}\rangle_{\lambda_{1},\lambda_{2}}|\leq1$, we obtain
		\begin{eqnarray}\label{I1s2a1}
		\qty| \ \Big\langle \mathcal{\tilde{A}}_{3,1}^{\prime}B_{3,1}\mathcal{\tilde{C}}_{3,1}^{\prime}\Big\rangle \ | \ &&\leq\iint  d\lambda_{1} \ d\lambda_{2} \ \mu (\lambda_{1})  \ \mu (\lambda_{2})  \ \big| \langle A_{3,1}\rangle_{\lambda_{1}}+\langle A_{3,2}\rangle_{\lambda_{1}} \big|  \ \big| \langle C_{3,1}\rangle_{\lambda_{2}}+\langle C_{3,2}\rangle_{\lambda_{2}}+\langle C_{3,3}\rangle_{\lambda_{2}}\big|
		\end{eqnarray}
		
The other terms can also be written in a similar manner as Eq. (\ref{I1s2a1}). Thus, from Eq. (\ref{n31s2a1}), we obtain the following
		\begin{equation}\label{S3bs2}
			(\mathcal{B}_{3}^{\prime})_{bl}\leq \left(\iint d\lambda_{1}\ d\lambda_{2} \ \mu(\lambda_{1}) \  \mu(\lambda_{2}) \ \delta_{1}^{\prime} \ \delta_{2}^{\prime}\right)^{\frac{1}{2}} 
		\end{equation}
		where  $\delta_{1}^{\prime}=\big[|\langle{A_{3,1}}\rangle_{\lambda_{1}}+\langle{A_{3,2}}\rangle_{\lambda_{1}}|+|\langle{A_{3,2}}\rangle_{\lambda_{1}}+\langle{A_{3,3}}\rangle_{\lambda_{1}}|+|\langle{A_{3,3}}\rangle_{\lambda_{1}}+\langle{A_{3,4}}|+|\langle{A_{3,4}}\rangle_{\lambda_{1}}-\langle{A_{3,1}}\rangle_{\lambda_1}|\big]$ and $\delta_{2}^{\prime}=\big[|\langle{C_{3,1}}\rangle_{\lambda_{2}}+\langle{C_{3,2}}\rangle_{\lambda_{2}}+\langle{C_{3,3}}\rangle_{\lambda_{2}}|+|\langle{C_{3,1}}\rangle_{\lambda_{2}}+\langle{C_{3,2}}\rangle_{\lambda_{2}}-\langle{C_{3,3}}\rangle_{\lambda_{2}}|+|\langle{C_{3,1}}\rangle_{\lambda_{2}}-\langle{C_{3,2}}\rangle_{\lambda_{2}}+\langle{C_{3,3}}\rangle_{\lambda_{2}}|+|-\langle{C_{3,1}}\rangle_{\lambda_{2}}+\langle{C_{3,2}}\rangle_{\lambda_{2}}+\langle{C_{3,3}}\rangle_{\lambda_{2}}| \big]$. 
		
		Since all the observables are dichotomic with eigenvalues $\pm1$, it is straightforward to derive that $\delta_{1}^{\prime}\leq6$ and $\delta_{2}^{\prime}\leq 6$. Therefore, from Eq.~(\ref{S3bs2}), integrating over $\lambda_{1}$ and $\lambda_{2}$ we obtain 
		\begin{equation}\label{S3fa}
			(\mathcal{B}_{3}^{\prime})_{bl}\leq 6
		\end{equation}
		
		\subsection{Optimal quantum bound of the asymmetric bilocality inequality for scenario:II}\label{aps22}
		Let us consider a suitable positive semi-definite operator $\gamma_{3}^{\prime}$ satisfying $\langle\gamma_{3}^{\prime}\rangle=\beta_{3}^{\prime}-\left(\mathcal{B}_{3}^{\prime}\right)_{Q}$. The existence of such operator is constructed by considering a set of operators $M^{\prime}_{3,j} \ \forall j\in \{1,2,3,4\}$ such that
		\begin{equation}\label{gamma3s2}
			\langle\gamma_{3}^{\prime}\rangle=\sum\limits_{j=1}^4\dfrac{\sqrt{\omega_{3,j}}}{2}\qty|M^{\prime}_{3,j}\ket{\psi}|^2
		\end{equation}
		where $\omega_{3,j}\geq0$ and $\omega_{3,j}=\omega^{A}_{3,j}\cdot\omega^{C}_{3,j}$. We choose $M^{\prime}_{3,j}$ and the quantity $\omega_{3,j}$ as the following way
		
		\begin{eqnarray}
			&&\qty|M^{\prime}_{3,j}\ket{\psi}|=\sqrt{\qty|\qty(\frac{\mathcal{\tilde{A}}_{3,j}^{\prime}}{\omega_{3,j}^A}\otimes\frac{\mathcal{\tilde{C}}_{3,j}^{\prime}}{\omega_{3,j}^C})\ket{\psi}|}-\sqrt{|B_{3,j}\ket{\psi}|} \ \  \ \forall j \in \{1,2,3,4\} \label{l31s2bl} \\
			&& \hspace{1cm} \omega^{A}_{3,j}=||\mathcal{\tilde{A}}_{3,j}^{\prime} \ket{\psi}||_{2} \ ;\ \omega^{C}_{3,j}=||\mathcal{\tilde{C}}_{3,j}^{\prime} \ket{\psi}||_{2} \label{w2s}
		\end{eqnarray}
		
		Putting $\qty|M^{\prime}_{3,j}\ket{\psi}|$ and $\omega_{3,j}^k$ from Eq.~(\ref{l31s2bl}) into Eq. (\ref{gamma3s2}) and, by using the inequalities given in footnotes (\ref{f2}) and (\ref{f3}), we obtain the quantum optimal value as follows
		\begin{eqnarray}\label{bs2}
			\left(\mathcal{B}_{3}^{\prime}\right)_{Q}^{opt} = \max \left[\prod_{k=A,C}\left(4\sum_{j=1}^{4}\left( \omega^{k}_{3,j}\right)^2\right)\right] ^{\frac{1}{4}} \ ; \ \ \ \ 	\text{with the optimality condition: }M^{\prime}_{3,j} \ \ket{\psi}=0 \ \ \forall j \in \{1,2,3,4\}
		\end{eqnarray}
		
		In the following we evaluate $\sum_{j=1}^{4}\left( \omega^{A}_{3,j}\right)^2$ and  $\sum_{j=1}^{4}\left( \omega^{C}_{3,j}\right)^2$ from Eq. (\ref{w2s}) separately.\\
		
		\paragraph{\textit{Evaluation of $\sum_{j=1}^{4}\left( \omega^{A}_{3,j}\right)^2$:}}\label{ps2}
		
		From Eq (\ref{w2s}) we obtain
		\begin{eqnarray}\label{w24ik1}
			\sum\limits_{j=1}^{4}\bigg(\omega^{A}_{4,j}\bigg)^2&=&\bra{\psi}(8+\{A_{4,2},(A_{4,1}+A_{4,3})\}+\{A_{4,4},(A_{4,3}-A_{4,1})\})\ket{\psi}
		\end{eqnarray}
		Note that in the above Eq. (\ref{w24ik1}) the quantities $A_{4,2}$ and $A_{4,4}$ appeared independently with $\{A_{4,2},(A_{4,1}+A_{4,3})\}$ and $\{A_{4,4},(A_{4,3}-A_{4,1})\}$ respectively. Thus, we can always define $A_{4,2}$ and $A_{4,4}$ independently. Hence, without loss of generality, we chose $A_{4,2}=(A_{4,3}+A_{4,1})/\nu_1$ and $A_{4,4}=(A_{4,3}-A_{4,1})/\nu_2$, where $\nu_1=|| \ (A_{4,3}+A_{4,1}) \ ||_2$ and $\nu_2=|| \ (A_{4,3}-A_{4,1}) \ ||_2$. therefore, the above Eq.~(\ref{w24ik1}) reduces to
		
		\begin{equation}
			\label{w24ik2}
			\sum\limits_{j=1}^{4}\bigg(\omega^{A}_{4,j}\bigg)^2=8+2\Bigg[\sqrt{4+2\sqrt{4-\langle\{A_{4,1},A_{4,3}\}\rangle}}\Bigg] \ \leq \  4 \left( 2+\sqrt{2}\right)   
		\end{equation}
		
		The maximum value of $\sum\limits_{j=1}^{4}\bigg(\omega^{A}_{4,j}\bigg)^2$ is then achieved when $\{A_{4,1},A_{4,3}\}=0$ which automatically implies $\nu_1=\nu_2=\sqrt{2}$. Therefore, for the optimal quantum violation, the linear constraints on Alice's observables are given as
		\begin{eqnarray}
			A_{4,1}-\sqrt{2}A_{4,2}+A_{4,3}=0  \ ; \ \ \ \ -A_{4,1}+A_{4,3}-\sqrt{2}A_{4,4}=0 \label{a44}
		\end{eqnarray}
		
		\paragraph{\textit{Evaluation of $\sum_{j=1}^{3}\left( \omega^{C}_{3,j}\right)^2$:}}\label{ps2b}
		By Eq. (\ref{w2s}), a straightforward calculation leads to
		\begin{equation}\label{w2s2}
			\sum_{j=1}^{4}\left( \omega^{C}_{3,j}\right)^2 \leq 12
		\end{equation}
		with the constraints on Charlie's observables are given by
		\begin{eqnarray}
			\{C_{3,1},C_{3,2}\} = \{C_{3,2},C_{3,3}\}=\{C_{3,1},C_{3,3}\}=0\label{cs2cond}
		\end{eqnarray}
		Finally, we obtain the optimal quantum value from Eqs. (\ref{bs2}), (\ref{w24ik2}) and (\ref{w2s2}) as 
		\begin{eqnarray}
			\left(\mathcal{\mathcal{B}}_{3}^{\prime}\right)_{Q}^{opt}=4\left[3\left(2+\sqrt{2}\right)\right]^{\frac{1}{4}}
		\end{eqnarray}

		\subsection{The state and observables for the optimal quantum violation of $(\mathcal{B}^{\prime})_{3}$}\label{aps23}
		
		We further obtain relationships between the observables of all the parties for achieving the optimal quantum violation in terms of the anti-commuting relations. The anti-commutation relations for Charlie's observables are already given in Eq.~(\ref{v32conds1}). From Eqs.~(\ref{condAlices1}), we obtain the following anti-commuting relations for Alice's observables as
		
		\begin{eqnarray}
			&&\{A_{3,1},A_{3,2}\}=\{A_{3,2},A_{3,3}\}=\{A_{3,3},A_{3,4}\}=-\{A_{3,1},A_{3,4}\}=\sqrt{2} \  \mathbb{I}_2 \ ; \ \ \ \{A_{3,1},A_{3,3}\}=\{A_{3,2},A_{3,4}\}=0  \label{as2cond}
		\end{eqnarray}
		
		By using the above relations between the observables  given by Eqs.~(\ref{cs2cond}) and (\ref{as2cond}) on the observables, one can always construct a set of observables for Alice and Charlie in the Hilbert space dimension $\mathcal{H}^d , \ \ \forall d\geq2$.

		Next, we recall the optimization condition obtained in the SOS method from the Eq.~(\ref{bs2}) to find the constraints on Bob's observable. The specific condition $M^{\prime}_{3,j} \ket{\psi} =0 \ \ \forall j\in\{1,2,3\}$ implies the following
		\begin{eqnarray} \label{bobcond3s2}
			B_{3,j}=\frac{\mathcal{\tilde{A}}_{3,j}^{\prime}}{\omega_{3,j}^A}\otimes\frac{\mathcal{\tilde{C}}_{3,j}^{\prime}}{\omega_{3,j}^C} \ \ \forall j \in \{1,2,3,4\}
		\end{eqnarray}
		
		We explicitly construct a set of observables for the Hilbert space dimension $\mathcal{H}^2$ are the following.
		\begin{eqnarray}\label{obss2}
			A_{3,1}&=& r\sigma_{x} + \sqrt{1-r^2} \ \sigma_{z} \ \ ; \ \ A_{3,2}= t\sigma_{x} + \sqrt{1-t^2} \ \sigma_{z}\ \ ; \ A_{3,3}=t\ \sigma_{x} -\sqrt{1-t^2}\ \sigma_{z} \ \ ; \ \ A_{3,4}= r \ \sigma_{x} - \sqrt{1-r^2} \ \sigma_{z}\nonumber \ \  \\
			C_{3,1}&=&\sigma_{x} \ \ ; \ \  C_{3,2}=\sigma_{y} \ \ ; \ \ C_{3,3}=\sigma_{z} \ \ \ \ \ \ [\text{where $r=\frac{1}{2}\sqrt{2-\sqrt{2}}$ and $t=\frac{1}{2}\sqrt{2+\sqrt{2}}$.}]
		\end{eqnarray}
		Bob's observables can be constructed from Eq. (\ref{bobcond3s2}). 
		Note that employing the above-mentioned observables, we find that the quantum optimal value $\left(\mathcal{B}_{3}^{\prime}\right)_{Q}^{opt}=6$ is achieved when two maximally entangled two-qubit states are shared between Alice-Bob and Bob-Charlie.\\
		
		\section{Detailed calculation for the asymmetric trilcality network scenario-II}\label{apt2}
		
		\subsection{Trilocal bound for the asymmetric Trilocality network scenario-II}\label{apt21}
		\begin{equation}
			\label{t3s2a2}
			\mathcal{T}_{3}^{\prime}=\sum_{j=1}^{3}|\mathcal{J}_{3,j}^{\prime}|^{\frac{1}{3}}\leq (\mathcal{T}_{3}^{\prime})_{tl} \ \ \ \text{with} \ \ \ \mathcal{J}_{3,j}^{\prime}=\Big\langle \mathscr{\tilde{A}}_{3,j}^{\prime} B_{3,j}\mathscr{\tilde{C}}_{3,j}^{\prime} \mathscr{\tilde{D}}_{3,j}^{\prime}\Big\rangle
		\end{equation}
		where the quantities $\mathscr{\tilde{A}}_{3,j}^{\prime}$, $\mathscr{\tilde{C}}_{3,j}^{\prime}$ and $\mathscr{\tilde{D}}_{3,j}^{\prime}$ are defined in Sec.~(\ref{secVI}) of the main text. Now, invoking the reproducibility condition given by Eq.~(\ref{blom}) and Since $\qty|\braket{B_{3,1}}_{\lambda_{1},\lambda_{2},\lambda_{3}}|\leq1$, we obtain
		\begin{eqnarray}\label{Jt3s2a2t2}
			\mathcal{J}_{3,1}^{\prime}&=&\iiint d\lambda_{1}  d\lambda_{2}  d\lambda_{3} \ \mu (\lambda_{1})  \mu (\lambda_{2}) \mu (\lambda_{3}) \ \Big|\langle A_{3,1}\rangle_{\lambda_{1}}+\langle A_{3,2}\rangle_{\lambda_{1}}+\langle A_{3,3}\rangle_{\lambda_{1}}-\langle A_{3,4}\rangle_{\lambda_{1}}\Big|  \ \Big| \langle C_{3,1}\rangle_{\lambda_{2}}+\langle C_{3,2}\rangle_{\lambda_{2}}\Big| \ \Big| \langle D_{3,1}\rangle_{\lambda_{3}}+\langle D_{3,2}\rangle_{\lambda_{3}}+\langle D_{3,3}\rangle_{\lambda_{3}}\Big|\nonumber\\
		\end{eqnarray}
		
		The terms $|\mathcal{J}_{3,2}^{\prime}|$, and $|\mathcal{J}_{3,3}^{\prime}|$ given by Eq. (\ref{t3s2a2}) can also be written in a similar manner as Eq. (\ref{Jt3s2a2t2}). Then, we obtain the following
		
		\begin{equation}\label{t3p}
			(\mathcal{T}_{3}^{\prime})_{bl}\leq \left(\iiint d\lambda_{1} d\lambda_{2} d\lambda_{3} \ \mu(\lambda_{1}) \mu(\lambda_{2}) \mu(\lambda_{3}) \ \  \eta_{1}^{\prime} \eta_{2}^{\prime} \eta_{3}^{\prime}\right)^{\frac{1}{3}}
		\end{equation}
		where  $\eta_{1}^{\prime}=\big[|\langle{A_{3,1}}\rangle_{\lambda_{1}}+\langle{A_{3,2}}\rangle_{\lambda_{1}}+\langle{A_{3,3}}\rangle_{\lambda_{1}}- \langle{A_{3,4}}\rangle_{\lambda_{1}}|+|\langle{A_{3,1}}\rangle_{\lambda_{1}}+\langle{A_{3,2}}\rangle_{\lambda_{1}}-\langle{A_{3,3}}\rangle_{\lambda_{1}}+ \langle{A_{3,4}}\rangle_{\lambda_{1}}|+|\langle{A_{3,1}}\rangle_{\lambda_{1}}-\langle{A_{3,2}}\rangle_{\lambda_{1}}+\langle{A_{3,3}}\rangle_{\lambda_{1}}+\langle{A_{3,4}}\rangle_{\lambda_{1}}|\big]$, \  $\eta_{2}^{\prime}=\big[|\langle{C_{3,1}}\rangle_{\lambda_{2}}+\langle{C_{3,2}}\rangle_{\lambda_{2}}|+|\langle{C_{3,2}}\rangle_{\lambda_{2}}+\langle{C_{3,3}}\rangle_{\lambda_{2}}|+|\langle{C_{3,3}}\rangle_{\lambda_{2}}-\langle{C_{3,1}}\rangle_{\lambda_{2}}| \big]$, and $\eta_{3}^{\prime}=\big[|\langle D_{3,1}\rangle_{\lambda_{3}}+\langle D_{3,2}\rangle_{\lambda_{3}}+\langle D_{3,3}\rangle_{\lambda_{3}}|+|\langle D_{3,1}\rangle_{\lambda_{3}}+\langle D_{3,2}\rangle_{\lambda_{3}}-\langle D_{3,3}\rangle_{\lambda_{3}}|+|\langle D_{3,1}\rangle_{\lambda_{3}}-\langle D_{3,2}\rangle_{\lambda_{3}}-\langle D_{3,3}\rangle_{\lambda_{3}}|\big]$. 
		
		Since all the observables are dichotomic with eigenvalues $\pm1$, it is straightforward to derive that $\eta_{1}^{\prime}\leq6$, $\eta_{2}^{\prime}\leq 4$, and $\eta_{3}^{\prime}\leq 5$. Therefore, from Eq.~(\ref{t3p}), integrating over $\lambda_{1}$ and $\lambda_{2}$ we obtain
		
		\begin{equation}\label{S3fb}
			(\mathcal{T}_{3})_{bl}\leq 2(15)^{\frac{1}{3}} \approx 4.93
		\end{equation}

		\subsection{Optimal quantum bound of the asymmetric trilocality inequality for scenario-II}\label{apt22}	
		To derive the optimal quantum bound of $\mathcal{T}_{3}^{\prime}$ without assuming the dimension of the system,  we again invoke the SOS approach discussed in the preceding Sec.~\ref{sosfors2}. Following the similar argument presented earlier in the Sec.~\ref{sosfors2}, we first show that there exists a positive semi-definite operator $\langle\Gamma_{3}^{\prime}\rangle=\zeta_3^{\prime}-\qty(\mathcal{T}_{3}^{\prime})_{Q}$. The existence of such operator can be proved by considering a set of operators $L_{3,j}^{\prime} \ \forall j\in \{1,2,3\}$ such that
		\begin{equation}
			\label{gamma3}
			\langle\Gamma_{3}^{\prime}\rangle=\sum\limits_{j=1}^4\dfrac{\qty(\omega_{3,j})^{\frac{1}{3}}}{2} \qty|L_{3,j}^{\prime}\ket{\psi}|^2
		\end{equation}
		where $\omega_{3,j}\geq0$ and $\omega_{3,j}=\omega^{A}_{3,j}\cdot\omega^{C}_{3,j}\cdot\omega^{D}_{3,j}$. We choose $L^{\prime}_{3,j}$ and the quantity $\omega_{3,j}$ as the following way
		
		\begin{eqnarray} \label{l31s2}
			&&\qty|L^{\prime}_{3,j}\ket{\psi}|=\qty|\qty(\frac{\mathscr{\tilde{A}}_{3,j}^{\prime}}{\omega_{3,j}^A}\otimes\frac{\mathscr{\tilde{C}}_{3,j}^{\prime}}{\omega_{3,j}^C} \otimes\frac{\mathscr{\tilde{D}}_{3,j}^{\prime}}{\omega_{3,j}^C})\ket{\psi}|^\frac{1}{3}-\qty|B_{3,j}\ket{\psi}|^{\frac{1}{3}} \ \  \ \forall j \in \{1,2,3\} \label{t31t2s} \\
			&& \hspace{1cm} \omega^{A}_{3,j}=||\mathscr{\tilde{A}}_{3,j}^{\prime} \ket{\psi}||_{2} \ ;\ \omega^{C}_{3,j}=||\mathscr{\tilde{C}}_{3,j}^{\prime} \ket{\psi}||_{2} \ ;\ \omega^{D}_{3,j}=||\mathscr{\tilde{D}}_{3,j}^{\prime} \ket{\psi}||_{2} \label{wdds}
		\end{eqnarray}

		Where $||  \cdot ||_2$ denotes the Frobenious norm given by $|| \ \mathcal{O} \ \ ||_2=\sqrt{\bra{\psi} \mathcal{O}^\dagger\mathcal{O}\ket{\psi}}$.
		
		Putting $L_{3,j}^{\prime}$ and $\omega_{3,j}$ from Eqs.~(\ref{t31t2s}) and (\ref{wdds}) into Eq. (\ref{gamma3}), and by using the inequalities given in footnotes (\ref{f2}) and (\ref{f3}), we obtain the quantum optimal value as follows
		\begin{eqnarray}\label{ts2}
			\left(\mathcal{T}_{3}^{\prime}\right)_{Q}^{opt} = \max \left[\prod_{k=A,C,D}\left(3\sum_{j=1}^{3}\left( \omega^{k}_{3,j}\right)^2\right)\right] ^{\frac{1}{6}} \ ; \ \ \ \ 	\text{with the optimality condition: } L^{\prime}_{3,j} \ \ket{\psi}=0 \ \ \forall j \in \{1,2,3,4\}
		\end{eqnarray}
		
Note that from Eqs.~(\ref{omega3as1}) and (\ref{v32}) $\max \sum_{j=1}^{3}\left( \omega^{A}_{3,j}\right)^2 = 16$	and $\max \sum_{j=1}^{3}\left( \omega^{C}_{3,j}\right)^2=9$, respectively. We evaluate  $\sum_{j=1}^{3}\left( \omega^{D}_{3,j}\right)^2$ as follows
		\begin{equation}\label{v32n}
			\sum_{j=1}^{3}\qty( \omega^{D}_{3,j})^2= \bra{\psi}(9+ \{D_{3,1},(D_{3,2}-D_{3,3})\}+\{D_{3,2},D_{3,3} \})\ket{\psi} =\bra{\psi}(9+ 3\mathbb{I}-(D_{3,1}-D_{3,2}+D_{3,3})^2)\ket{\psi} \leq 12\\
		\end{equation}
	
The above Eq. (\ref{v32n}) provides maximum value when 
		\begin{equation} \label{v32cond}
			D_{3,1}-D_{3,2}+D_{3,3}=0
		\end{equation}
		
		By placing the value of $\sum_{j=1}^{3}\qty(\omega^{A}_{3,j})^2$, $\sum_{j=1}^{3}\qty( \omega^{C}_{3,j})^2$, and $\sum_{j=1}^{3}\qty( \omega^{D}_{3,j})^2$ in Eq. (\ref{ts2}) we obtain the optimal quantum bound as
		\begin{eqnarray}
			\left(\mathcal{\mathcal{T}}_{3}^{\prime}\right)_{Q}^{opt}=6
		\end{eqnarray} 
		It is important to remark here that the optimal quantum value $\left(\mathcal{T}_{3}\right)_{Q}^{opt}=6$ is evaluated without specifying the dimension of both the system and observables. The optimal value fixes the states, and the observables are the following. 
		
		\subsection{The state and observables for the optimal quantum violation of $\mathcal{T}^{\prime}_{3}$}\label{apt23}
		
		We further obtain relationships between the observables of all the parties for achieving the optimal quantum violation. It follows from the earlier derived results (Eqs. \ref{condAlices1} and \ref{v32conds1}) and Eq. (\ref{v32cond}) the following anti-commuting relations of the observables for all the parties. 
		\begin{eqnarray}
			&&\{A_{3,1},A_{3,2}\}=\{A_{3,1},A_{3,3}\}=\{A_{3,1},A_{3,4}\}=\frac{2}{3}\mathbb{I}_{d} ; \ \{A_{3,2},A_{3,3}\}=\{A_{3,2},A_{3,4}\}=\{A_{3,3},A_{3,4}\}=-\frac{2}{3}\mathbb{I}_{d} \label{at2cond}\\
			&&\{C_{3,1},C_{3,2} \}=\{C_{3,2},C_{3,3} \}=-\{C_{3,1},C_{3,3}\} = \{D_{3,1},C_{3,2} \}=\{D_{3,2},D_{3,3} \}=-\{D_{3,1},D_{3,3}\}=\mathbb{I}_{d} \label{ct2cond}
		\end{eqnarray}
		
		By using the above relations between the observables  given by Eqs.~(\ref{at2cond}-\ref{ct2cond}) on the observables, one can always construct a set of observables for Alice and Charlie in the Hilbert space dimension $\mathcal{H}^d , \ \ \forall d\geq2$.

		Next, we recall the optimization condition obtained in the SOS method from the Eq.~(\ref{ts2}) to find the constraints on Bob's observable. The specific condition $L^{\prime}_{3,j} \ket{\psi} =0 \ \ \forall j\in\{1,2,3\}$ implies the following
		\begin{eqnarray}\label{t31t2}
			B_{3,j}=\frac{\mathscr{\tilde{A}}_{3,j}^{\prime}}{\omega_{3,j}^A}\otimes\frac{\mathscr{\tilde{C}}_{3,j}^{\prime}}{\omega_{3,j}^C} \otimes\frac{\mathscr{\tilde{D}}_{3,j}^{\prime}}{\omega_{3,j}^C}
		\end{eqnarray}
		We explicitly construct a set of observables for the Hilbert space dimension $\mathcal{H}^2$ are the following.
		\begin{eqnarray}\label{ts2obs}
			A_{3,1}&=& \frac{\sigma_{x}+\sigma_{y}+\sigma_{z}}{\sqrt{3}} \ ;\              A_{3,2}=\frac{\sigma_{x}+\sigma_{y}-\sigma_{z}}{\sqrt{3}} ; \ 
			A_{3,3}= \frac{\sigma_{x}-\sigma_{y}+\sigma_{z}}{\sqrt{3}};\  A_{3,4}=\frac{-\sigma_{x}+\sigma_{y}+\sigma_{z}}{\sqrt{3}} \ ; \ C_{3,1}=\sigma_{z} \nonumber\\ 
			C_{3,2}&=&\left(\frac{\sqrt{3}}{2}\sigma_{x}+\frac{\sigma_{z}}{2}\right); \ C_{3,3}=\left(\frac{\sqrt{3}}{2}\sigma_{x}-\frac{\sigma_{z}}{2}\right) \ ; \ D_{3,3}=-\sigma_{z} \ ; \ D_{3,1}=\left(\frac{-\sqrt{3}}{2}\sigma_{x}+\frac{\sigma_{z}}{2}\right); \  D_{3,2}=\left(-\frac{\sqrt{3}}{2}\sigma_{x}-\frac{\sigma_{z}}{2}\right)
		\end{eqnarray}
		
		Note that Bobs observables can be constructed from Eq. (\ref{t31t2}). Now, employing the above-mentioned observables, we find that the quantum optimal value $\left(\mathcal{T}_{3}^{\prime}\right)_{Q}^{opt}=6$ is achieved when three maximally entangled two-qubit states are shared between Alice-Bob, Charlie-Bob and Diana-Bob.
		
\end{widetext}
	
\bibliography{Asymmetric_Network}

\end{document}